\pdfoutput=1

\documentclass{aa} 

\usepackage{txfonts}
\usepackage{graphicx}
\usepackage{natbib}

\newcommand{\FIG}[1]{#1}

\bibpunct{(}{)}{;}{a}{}{,}
\def\lapp{\mathbin{\raise2pt \hbox{$<$} \hskip-9pt \lower4pt \hbox{$\sim$}}}
\def\gapp{\mathbin{\raise2pt \hbox{$>$} \hskip-9pt \lower4pt \hbox{$\sim$}}}
\def\bfv{\mbox{\bf v}}
\def\bfx{\mbox{\bf x}}
\def\bfm{\mbox{\bf m}}
\def\bfB{\mbox{\bf B}}
\def\bfE{\mbox{\bf E}}
\def\bfI{\mbox{\bf I}}
\def\bfSg{\mbox{\bf S}_g}
\def\bfSt{\mbox{\bf S}}
\def\bfS{\mbox{\bf S}_{\rm em}}
\begin{document}
\title{Extragalactic jets with helical magnetic fields: relativistic MHD simulations}

  \titlerunning{Extragalactic relativistic jets}
  \authorrunning{Keppens et al.}
  \author{R. Keppens
           \inst{1,2,3}
          \and Z. Meliani
           \inst{2}
           \and  B. van der Holst\thanks{\emph{Present adress:} University of Michigan, Space physics research laboratory, 2455 Hayward
Ann Arbor, Michigan 48109-2143, USA.}
           \inst{1}
           \and  F. Casse
           \inst{4}
          }

   \offprints{R. Keppens}

\institute{Centre for Plasma Astrophysics, K.U.Leuven (Leuven Mathematical Modeling and Computational Science Center), Celestijnenlaan 200B,
3001 Heverlee, Belgium.
\and FOM-Institute for Plasma Physics Rijnhuizen, Nieuwegein
\and Astronomical Institute, Utrecht University\\
\email{Rony.Keppens@wis.kuleuven.be}
\and AstroParticule \& Cosmologie (APC),
      Universit\'e Paris Diderot, 10 rue Alice Domon et L\'eonie Duquet,
75205 Paris Cedex 13, France\\
 \email{fcasse@apc.univ-paris7.fr}
 }

   \date{Received ... / accepted ...}

 \abstract{
Extragalactic jets are inferred to harbor dynamically important, organized magnetic fields which presumably aid in the collimation of the relativistic jet flows. We here explore by means of grid-adaptive, high resolution numerical simulations the morphology of AGN jets pervaded by helical field and flow topologies. 
We concentrate on morphological features of the bow shock and the jet beam behind the 
Mach disk, for various jet Lorentz factors and magnetic field helicities.}
{
We investigate the influence of helical magnetic fields on
jet beam propagation in overdense external medium.
We adopt a special relativistic magnetohydrodynamic (MHD) viewpoint on the
shock-dominated AGN jet evolution. Due to the Adaptive Mesh Refinement (AMR), 
we can concentrate on the long term evolution of
kinetic energy dominated jets, with beam-averaged Lorentz factor $\Gamma\simeq 7$, as they penetrate into denser clouds. These jets have near-equipartition 
magnetic fields (with the thermal energy), and radially varying 
$\Gamma(R)$ profiles within the jet radius $R<R_j$ maximally reaching  $\Gamma\sim 22$.}
{
We use the AMRVAC code, employing a novel hybrid block-based AMR strategy, to
compute ideal plasma dynamics in special relativity. We combine this
with a robust second-order shock-capturing scheme and a
diffusive approach for controlling magnetic monopole errors.}
{
We find that the propagation speed of the bow shock systematically exceeds the
value expected from estimates using beam-average parameters, in accord with the centrally peaked $\Gamma(R)$ variation. The helicity of the beam magnetic field is effectively transported down the beam, with compression zones in between diagonal internal cross-shocks showing stronger toroidal field regions. In comparison with
equivalent low-relativistic jets ($\Gamma\simeq 1.15$) which get surrounded by
cocoons with vortical backflows filled by mainly toroidal field, the high speed
jets demonstrate only localized, strong toroidal field zones within the 
backflow vortical structures. The latter are ring-like due to our
axisymmetry assumption and may further cascade to smallscale in 3D. We find evidence for
a more poloidal, straight field layer, compressed between jet beam and 
backflows. This layer decreases the destabilizing influence of the backflow on the jet beam.
In all cases, the jet beam contains rich cross-shock patterns, across which part of the kinetic energy gets transferred.
For the high speed reference jet considered here, significant jet deceleration only occurs beyond distances exceeding ${\cal O}(100 R_j)$, as the axial flow can reaccelerate downstream to the internal cross-shocks.
This reacceleration is magnetically aided, due to field compression across the internal shocks which pinch
the flow.
}
{
}

\maketitle

\section{Motivation}

Relativistic jets represent extremely energetic phenomena in
astrophysics. They are associated with (1) a variety of compact objects, 
(2) with Active Galactic Nuclei (AGN) carrying energy fluxes of 
$10^{45} - 10^{48}{\rm ergs/s}$ \citep{Celottietal97, Tavecchioetal00}, 
or (3) with micro-quasar systems (energy flux of $10^{35} {\rm ergs/s}$). 
These high energies are somehow extracted from the inner part of the system 
and this energy is transported over long distances by means of a detectable 
collimated jet. A large amount of this energy gets deposited by the jet in 
the surrounding medium, as only a small fraction of the jet energy is 
dissipated in the innermost region \citep{Sambrunaetal06}. 
In many AGNs, the kinetic energy flux is 
comparable to the bolometric radiative luminosity of the central part
\citep{Rawlings&Saunders91, Xu&Livio&Baum99, Sambrunaetal06}. 
This implies that a study of the interaction of the magnetized jet with external medium 
could help provide model constraints for jet formation and collimation.
The jet is assumed to be powered by a spinning
black hole \citep{Blandford&Znajek77, Begelmanetal84}, and/or by the 
surrounding accretion disk corona \citep{Miller&Stone00}. 
The disk-jet is launched by general relativistic magnetohydrodynamic (GRMHD) 
mechanisms, and is accelerated to high Lorentz factor. 
The observations show that AGN jets propagate in the parsec scale with a Lorentz 
factor $\Gamma\sim 10- 30$ \citep{Kellermannetal04}, and in some AGN types
(blazar, QSO), the jets are
relativistic even at kpc scale with $\Gamma\sim 5- 30$ \citep{Tavecchioetal04}.
In such objects, the radio to X-ray observations cannot be explained by 
the presence of a unique synchrotron component. Scenarios involving synchrotron
and inverse Compton scattering of synchrotron photons are able to fit 
observations, under the assumption that the jet decelerates from the subpc scale
to the larger scale \citep{Georg03}. This can be achieved if an outer slower 
layer surrounds the central relativistic jet ``spine" \citep{Ghise05}, 
since continuous electron acceleration acts at the boundary layer between jet sheath and ``spine" \citep{Stawarz&Ostrowski02}. 
This kind of jet structure is supported by direct radio observations 
\citep{Giro04} and leads to radial jet structure where the Lorentz factor 
decreases radially.
Such a stratification of the jet 
where the Lorentz factor increases towards the axis, has been suggested from 
observations, both in microquasars \citep{meier03}, and in AGN \citep{Jester06, 
Dulwichetal07}. Such transverse jet stratification is also supported by 
models of jet launch scenarios \citep{Koideetal01, McKinney06, Melianietal06}. 
However, most numerical investigations to date have ignored radial stratification of the jet velocity, and have made
simplifying assumptions on the magnetic topology.
In this paper, we present numerical simulations of the propagation of magnetized, relativistic, radially stratified jets. These are important to
understand the impact of more realistic jet topologies on the surrounding medium, and how this
in turn helps to deduce properties of jet formation mechanisms. We look in
particular to the effects of helical magnetic fields on the relativistic
jet propagation in external media.  

The magnetic field seems to play a significant role in the jet collimation and 
contributes to its acceleration \citep{Lietal92, Contopoulos94, Fendt97, Koideetal01, Vlahakis&Konigl04, BogovalovTsinganos05, Melianietal06}. 
There is observational evidence for
an intrinsic magnetic field in relativistic jets. In BL-Lac objects, 
VLBI observations of the polarisation of the synchrotron emission, and especially the
rotation measure,
have shown that the magnetic field is systematically tilted from the jet axis
\citep{Asadaetal02, Gabuzdaetal04}, which indicates the presence of helical 
magnetic fields. 

Significant progress has been made 
regarding the numerical modeling of relativistic jet propagation, especially
for relativistic hydrodynamic models \citep{Hardeeetal05, Peruchoetal06}. Relativistic MHD studies of jet propagation have been
undertaken by several authors, mostly restricted to axisymmetric (or 2.5D) configurations. 
Recently, 3D simulations of magnetized `spine-sheath' jets have emerged as well \citep{mizuno07}, for purely axial magnetic field configurations at modest Lorentz factors of $\Gamma=2.5$. Combined with linear stability results for relativistic 
`top-hat' profiles (i.e. uniform jet bounded by a uniform sheath), these results confirm the possibility of jet stabilization by invoking radial structure. We will restrict
our study to axisymmetric configurations (as even our grid-adaptive simulations are still fairly computationally expensive for
long-term evolutions),  but emphasize high Lorentz factor flow regimes in helical field topologies. Original RMHD simulations were presented by~\citet{vanputten}, where
low Lorentz factors in toroidally magnetized jets were simulated. A significant step forward was presented by~\citet{komiss99}, 
with studies of light, Lorentz factor $\Gamma=10$ jets pervaded by purely 
toroidal magnetic fields. By confronting a Poynting
flux dominated, with a kinetic energy dominated jet, clear morphological 
differences were identified and explained: extensive cocoons
form in kinetic energy dominated cases, while `nose cones' develop with high 
stand-off distances between Mach disk and bow shock for
the Poynting dominated jets. The latter correlate to highly magnetized, 
shocked plasma, self-collimated by strong magnetic pinching.
These two distinctive cases have been used subsequently by various authors, also as benchmarks for the rapidly growing code parc
capable of performing special relativistic MHD computations \citep{kepmel07,mignone1}, while a first more comprehensive study,
including also purely poloidally magnetized jets was presented by~\citet{leis05}. The latter confirmed the purely toroidal field 
cases, but also demonstrated that purely poloidal field cases did not develop pronounced nose cones. The latter were found less
susceptible to internal shock deformations, due to the stabilizing magnetic tension. Hence, fairly smooth cocoons and stable beams
resulted. The authors speculated how the resulting brightness differences would allow to distinguish inherent magnetic topologies. We here revisit this suggestion, by making a more concentrated effort on kinetic energy dominated jets alone, with a more gradual change from
toroidal to near poloidal fields. The helical field configurations presented indeed confirm the earlier trends, but our grid-adaptive models
do show rich shock-structured beams also for more poloidal field regimes. 
The code used here is the AMRVAC code \citep{Keppens03,Melianietal07,holst07}, which has also been tested on a variety of stringent
1D and multi-D relativistic MHD problems \citep{Holst08}. 
In section~\ref{sec1}, we list the governing equations, Sect.~\ref{sec2} provides details on the numerical strategy, while
Sect.~\ref{sec3} discusses the main results.

\section{Magnetohydrodynamics in special relativity}\label{sec1}
In special relativistic theory, where material particles move through four-dimensional spacetime with speeds strictly less than the speed of light $c$, the governing conservation laws express particle number conservation, and energy-momentum conservation expressable as the vanishing divergence of a stress-energy tensor. This latter four-tensor includes 
(perfect) gas as well as electromagnetic contributions, and in general, the full dynamics governed by gas and electromagnetic field variations needs to solve the full set of Maxwell equations as well. When
we adopt the ideal MHD approximation, in which the electric field in the comoving frame vanishes identically, a numerically convenient set of conservation laws results when we
choose a `lab' Lorentzian reference frame with four-coordinates $(ct,\bfx)^T$, where $\bfx=(x_1,x_2,x_3)$ are the three spatial orthogonal coordinate axes. The spacetime
metric for special relativistic applications is the usual Minkowski metric $g^{\alpha\beta}={\rm diag}(-1,1,1,1)$, where the greek indices take values from $0,1,2,3$. If we indicate
the proper density as experienced in the local rest frame with $\rho=m_0 n_0$, where $m_0$ denotes the particle rest mass and $n_0$ the proper number density, Lorentz contraction
results in a lab frame number density $\Gamma n_0$, with the Lorentz factor $\Gamma=1/\sqrt(1-v^2/c^2)$. Particle number conservation as written for the fixed reference frame is then written using the variable $D=\Gamma\rho$ as
\begin{equation}
\frac{\partial D}{\partial t}+\nabla \cdot \left(D\bfv\right) = 0 \,.
\label{dens}
\end{equation}
This includes the three-velocity $\bfv$, whose magnitude $v<c$. The temporal component of the divergence of the stress-energy can be written in the same laboratory frame as
\begin{equation}
\frac{\partial \,}{\partial t} \left( \tau + D c^2 \right ) +\nabla \cdot \bfSt = 0 \,.
\label{energy}
\end{equation}
The energy flux $\bfSt$ expression will be given below, and the total energy in the lab frame is then split off in a rest mass contribution $D c^2$, and the partial energy $\tau$ made up of gas, magnetic and electric field energy densities from
\begin{equation}
\tau = \tau_{\rm g} + \frac{B^2}{2\mu_0}+\epsilon_0\frac{E^2}{2} \,. \label{tau}
\end{equation}
In Eq.~(\ref{tau}), $B$ and $E$ indicate the magnitude of the usual three-vector magnetic $\bfB$ and electric $\bfE$ fields, respectively. When we adopt a simplifying polytropic equation
of state where the specific internal energy of the gas is directly related to the proper (rest frame) density and pressure, i.e. $p/(\gamma -1)\rho$ with polytropic index $\gamma$, 
the gas contribution is
\begin{equation}
\tau_{\rm g}= \rho \left(c^2+\frac{\gamma p}{(\gamma-1)\rho}\right) \Gamma^2 - p - D c^2 \,.
\end{equation}
The term between brackets represents the relativistic specific enthalpy containing a rest mass contribution. The energy flux in Eq.~(\ref{energy}) can also be split in a gas and electromagnetic contribution as in $\bfSt=\bfSg+\bfS$, where
the latter are then quantified from
\begin{eqnarray}
\bfSg & = & \rho \left(c^2 +\frac{\gamma p}{(\gamma -1)\rho}\right) \Gamma^2 \bfv \,,\nonumber \\
\bfS & = & \frac{\bfE \times \bfB}{\mu_0} \,.
\end{eqnarray}
Obviously, $\bfS$ is the Poynting flux. With this notation, the spatial part of the stress-energy divergence, using the ideal MHD approximation, can be written in the lab frame as
\begin{equation}
\frac{\partial \bfSt }{\partial t}  +\nabla \cdot \left( 
\bfSt\bfv  + p_{\rm tot} c^2 \bfI 
-\frac{c^2}{\mu_0}\frac{\bfB\bfB}{\Gamma^2} -\frac{(\bfv\cdot\bfB)}{\mu_0} \bfv \bfB \right) 
= 0 \,. \label{mom}
\end{equation}
The $\bfI$ denotes the three by three identity matrix, and the total pressure $p_{\rm tot}$ is computed from
\begin{equation}
p_{\rm tot} = p + p_{\rm mag} = p + \frac{1}{2\mu_0} \left( \frac{\bfB\cdot\bfB}{\Gamma^2}+\frac{(\bfv\cdot\bfB)^2}{c^2} \right) \,.
\end{equation}
The system is then closed with the homogeneous Maxwell equations
\begin{eqnarray}
\nabla \cdot \bfB  & = & 0 \,, \nonumber \\
\frac{\partial \bfB}{\partial t} & = & \nabla \times ( \bfv \times \bfB ) \,.
\label{ind}
\end{eqnarray}
The latter is completely identical to the non-relativistic induction equation in ideal MHD formulations, since vanishing electric fields in the comoving frame implies similarly
$\bfE = - \bfv \times \bfB $. This allows to write the electric energy density
in Eq.~(\ref{tau}) as $\epsilon_0 E^2/2=\epsilon_0[B^2 v^2 - (\bfv\cdot\bfB)^2]/2$, and the
Poynting flux as $\bfS=[B^2\bfv -(\bfv\cdot\bfB)\bfB]/\mu_0$.

\section{Numerical approach}\label{sec2}
\subsection{Algorithmic details}
For the numerical solution of the relativistic MHD equations, we actually combine Eq.~(\ref{energy}) and the particle conservation law Eq.~(\ref{dens}) to obtain the following
conservation law
\begin{equation}
\frac{\partial \tau}{\partial t} +\nabla \cdot \left( (\tau+ p_{\rm tot})\bfv -\frac{(\bfv\cdot\bfB)}{\mu_0} \bfB \right) = 0 \,.
\end{equation}
It is then also necessary for the numerical approach to exploit a scaling where
$c=1$ and thus setting electromagnetic units where $\mu_0=1=\epsilon_0$. This
scaling will be exploited from here onwards. In terms of the conserved variables
$[D, \bfSt, \tau, \bfB]$ the usual non-relativistic ideal MHD equations in terms
of the conserved quantities $[\rho, \bfm\equiv \rho\bfv, e\equiv \rho v^2/2 + p/(\gamma-1)+ B^2/2, \bfB]$ are directly obtained from the 
limit $\Gamma\rightarrow 1$. In our conservation-law oriented integration 
strategy, the induction equation~(\ref{ind}) is written as
\begin{equation}
\frac{\partial \bfB}{\partial t} +\nabla \cdot \left( \bfB\bfv -\bfv\bfB \right) = 
\eta_d \nabla (\nabla \cdot \bfB) \,.
\end{equation}
The coefficient $\eta_d$ is chosen to correspond to the maximal
allowed diffusion coefficient which still complies with an unmodified
Courant-Friedrichs-Lewy constrained time step $\Delta t$. This acts to diffuse potential numerical monopole errors at their maximal rate, and has been used in various
non-relativistic MHD applications.
We refer to \citet{Keppens03}
for a comparitive study between this and other popular source term strategies
for $\nabla \cdot \bfB$ treatments in an AMR framework, and to \citet{holst07}
for its use in AMR in combination with curvilinear coordinates.

When we further introduce the auxiliary variable $\xi$ from
\begin{equation}
\xi= \Gamma^2 \left(\rho + \frac{\gamma p}{(\gamma-1)}\right) \,,
\end{equation}
the Lorentz factor in essence depends on $\Gamma(\bfSt,\bfB;\xi)$, as one can
write $\bfv=\left(\bfSt + \xi^{-1}(\bfSt\cdot\bfB)\bfB\right)/(\xi+B^2)$.
The defining relation~Eq.(\ref{tau}) then becomes 
\begin{equation}
\xi-\frac{\gamma-1}{\gamma}\frac{(\xi-\Gamma D)}{\Gamma^2}+B^2-\frac{1}{2}\left[ \frac{B^2}{\Gamma^2}+\frac{(\bfSt \cdot \bfB)^2}{\xi^2}\right]-\tau-D=0.
\end{equation}
A root-finding algorithm (Newton-Raphson) is implemented to compute $\xi$
from this expression. This then suffices to make the needed conversions from
conservative to primitive variables $[\rho, \bfv, p, \bfB]$. 

The shock-capturing conservative discretization (with the monopole source
term strictly speaking destroying perfect conservation for $\bfB$ only) is then
a TVDLF \citep{tothod} type method, in which only the maximal physical
propagation speed needs to be computed. This is done 
using a (slightly modified from Numerical Recipes) Laguerre's method to 
compute both speed pairs pertaining to slow and fast magneto-acoustic
waves as the roots of a quartic polynomial (see e.g.~\citet{zanna}). Since all 4 roots $\lambda$ must lie in the interval
$]-1,+1[$ but can become notoriously close to each other and unity, it helps to
transform~\citep{jeroen} to a variable $\mu=1/(1-\lambda)$ with well-seperated roots
on $]0.5,+\infty[$. The TVDLF scheme is used with a Hancock predictor, and
is second order accurate for smooth solutions. This implies limited linear constructions to obtain cell edge from cell center
quantities, and in this process we employ the set $[\rho, \Gamma\bfv, p, \bfB]$. 
\subsection{AMR strategy and numerical setup}
All jet simulations are done a domain size $[R,Z]\in[0,40]\times[0,200]$, and
allow for six grid levels (including the base level), achieving an effective 
resolution of $3200\times 8000$. The jet internal structure will be specified 
in detail below, and it initially occupies only the region 
$[0,R_{\rm j}]\times [0,Z_{\rm j}]$.
The exterior region will always represent a higher density
medium in our simulations and is meant to mimick a denser `cloud',
leading to jet deceleration by entrainment and dissipation, at 
least in ${kpc}$ scale. 
Our normalization always sets the jet radius $R_{\rm j}=1.5$ and height 
$Z_{\rm j}=3$. We will typically compute till times beyond $t=210$, and due to 
the adopted scaling where the light speed $c=1$, we thus follow jet propagation
for at least 70 light crossing times of the jet beam diameter $2R_{\rm j}$. 
To convert from dimensionless, computed values to physical quantities, one may adopt~\citep{harris}
a typical radius for an AGN jet of 0.05 $pc$, a cloud number density of 10 ${\rm cm}^{-3}$, and 
the light speed. 
Our AMR scheme exploits a Richardson type error estimator to dynamically create
or remove finer level grids where needed. In addition to this estimator,
we always enforce the highest grid level around this inlet corner, such that we
have 120 grid cells through the jet radius $R_{\rm j}$ and the same amount
of cells through the initial jet height $Z_{\rm j}$. In the Richardson 
process itself, two low-order $t^{n+1}$ solutions with coarsened grid 
spacing $2\Delta x$ are 
constructed from the known solutions at
resolution $\Delta x$ at times $t^n$ and $t^{n-1}$, by reversing the
order of the time integration and coarsening operations. When a weighted 
average of selected components exceeds a tolerance parameter $\epsilon_{\rm tol}=0.005$, new grids are created. In all runs, these selected components
include the lab `density' $D$, partial energy density $\tau$, and magnetic field
component $B_\varphi$, with weight ratios $2:1:1$.

The boundary conditions for the simulations enforce the
primitive variable profiles as discussed in the next section within $R<R_{\rm j}$ at the lower boundary. The remainder of this bottom $Z=0$ boundary is treated as a symmetry boundary for $D, S_R, S_\varphi, \tau$ and $B_Z$ while
we ensure the vanishing of $S_Z, B_R, B_\varphi$. This acts as a kind of
reflecting underlying `disk' configuration from which potential backflows
deflect. The lateral and top boundaries are open, while the symmetry axis
$R=0$ enforces the usual (a)symmetry conditions.

\section{AGN jet computations}\label{sec3}
\subsection{Jet inlet conditions}
In all 8 simulations discussed below, we fix the polytropic index $\gamma=5/3$.
In fact, all models investigated in this paper represent underdense jets,
and have typically maximum Lorentz factor at the axis of $22$. Then, we can expect from analogous 1D Riemann problems that the forward shock and reverse shock are both near Newtonian, so
that it is quite adequate to use this Newtonian polytropic index value. 
The magnetic
configuration in the lab frame is initially given by 
\begin{eqnarray}
B_R & = & 2 B_0 \frac{R_{\rm j}}{Z_{\rm j}} \frac{ \left(\frac{Z}{Z_{\rm j}}\right)^3 \tanh \left(\frac{Z}{Z_{\rm j}}\right)^4 \tanh \left(\frac{R}{R_{\rm j}}\right)^2}{\frac{R}{R_{\rm j}} \cosh \left(\frac{Z}{Z_{\rm j}}\right)^4} \nonumber, \\
B_Z & = & B_{\rm c} +  \frac{B_0}{\left[\cosh \left(\frac{R}{R_{\rm j}}\right)^2 \right]^2 \cosh \left(\frac{Z}{Z_{\rm j}}\right)^4} , \nonumber \\
B_\varphi & = & \left\{ \begin{array}{lr} B_1 \tanh \left(\frac{R}{a}\right) & \,\,\,\,{\rm in} \,\,\,\,\,\,\, [0,R_{\rm j}]\times [0,Z_{\rm j}], \\
0 & \,\,\,\, {\rm elsewhere}. \end{array} \right. 
\end{eqnarray}
This solenoidal field is fully helical internal to the jet whenever $B_1\ne 0$, while it is purely poloidal elsewhere.
The parameter ``$a$'' for the azimuthal field variation is held fixed at $a=5$, and in all but our purely toroidal field simulation (where the exterior is unmagnetized), the magnetic field
strength of the cloud $B_c=0.01$, corresponding to a weak uniform background magnetic field. We explore the relative importance of the jet
internal toroidal versus poloidal field components by varying the parameters 
$B_0$ and $B_1$. The proper densities of the jet and cloud are typically
fixed at $\rho_{j}=100$ and $\rho_c=1000$, except for one run where the 
cloud density $\rho_c$ is decreased twofold. Hence, we restrict the discussion to
the propagation of relativistic underdense jets, since they
yield deceleration of the jet and formation of the typically complex cocoons. Moreover, 
we explore only relativistic jets dominated by kinetic energy,
which are then characterised by a strong shock \citep{Appl&Camenzind88}, which in turn are
efficient in particle acceleration \citep{Begelman&Kirk90}.
The initial pressure distribution is computed from
\begin{equation}
p = p_{\rm j} + \frac{1}{2}B_0^2 - \frac{1}{2}\left(B_\varphi^2(R,Z)+B_Z^2(R,Z)\right)\, ,
\end{equation}
where the reference value for the parameter $p_{\rm j}=1$ (representing a cold jet).
Finally, the initial flow field $\bfv$ is quantified by the following prescription
\begin{eqnarray}
v_Z & = & \alpha \frac{B_\varphi}{\sqrt{\rho}(R/a)}, \nonumber \\
v_\varphi & = & \frac{B_\varphi}{\sqrt{\rho}}, \nonumber \\
v_R & = & 0 \,.
\end{eqnarray}
It vanishes outside the jet region $[0,R_{\rm j}]\times [0,Z_{\rm j}]$, and includes jet rotation whenever $B_1\ne 0$. 
Note that the requirements $p>0$ and $v<1$ restrict the possible choices for the parameters $p_{\rm j}$ and $\alpha$, once values for $B_0$, $B_1$ ($B_c$) and $\rho_{\rm j}$ are
adopted. This prescription of the
primitive variables $\rho$, $\bfv$, $p$ and $\bfB$ is inspired by previous non-relativistic ideal MHD models as exploited in~\citet{CasseMar}, and is such that it then ensures the
radial force balance along the lower boundary $Z=0$. We quantify in what follows how these parameters translate into various dynamically important dimensionless parameters
like Lorentz factor, Mach number, Alfv\'en Mach number, magnetization parameter, as averaged over the jet radius. In our implementation, we found it convenient to prescribe these 
jet profiles at the inlet boundary $Z=0$ directly from the above primitive variable expressions within those bottom ghost cells with cell centers $Z<0$ and $R<R_{\rm j}$.

In Table~\ref{tab1}, we list the input model parameters for the eight cases
studied. The first three runs only differ in the value of $\alpha$, and
mimick the effect of going from nearly non-relativistic speeds (NR), to
mildly relativistic (MR), to fully relativistic flows in our reference case 
(Ref1). The next three models (Pol, Tor, Ref2) explore variations in magnetic 
field
topology (from almost purely poloidal to fully toroidal). The final two cases
look at modest changes (factor 2) in external density (Ref3) or overall
pressure (Ref4). 
Note that for all the cases considered, the entire jet rotation profile enforced at the inlet $Z=0$ is within the light cylinder, i.e. within the jet radius $R<R_{\rm j}$, we have $\mid v_\varphi-v_Z B_\varphi/B_Z \mid $ increasing from zero to 0.63 for most of the models.
In Table~\ref{tab2}, we list the corresponding dimensionless parameters which
characterize the internal jet properties. We list the Lorentz factor $\Gamma$,
the Mach number $M=v_Z/c_s$ with the sound speed given by
$c_s=\sqrt{\gamma(\gamma-1) p/((\gamma-1)\rho+\gamma p)}$, and the relativistic proper Mach number ${\cal M}=M\Gamma/\Gamma_{s}$ where $\Gamma_s=1/\sqrt(1-c_s^2)$ is the Lorentz factor computed from the sound speed. We further give the
Alfv\'en Mach number $M_a=v_Z/c_a$ where the Alfv\'en speed is found from 
\citep{Lichnerowicz67}
\[ c_a=\sqrt{\frac{\sigma}{\sigma+1+\frac{\gamma p}{(\gamma-1)\rho}}} . \]
The latter contains the ratio of magnetic to rest mass energy density
\begin{equation}
\sigma=\frac{2 p_{\rm mag}}{\rho} , \label{sigma}
\end{equation}
which together with the reciprocal plasma-$\beta$ parameter
\begin{equation}
\beta_r=\frac{p_{\rm mag}}{p} , \label{betar}
\end{equation}
is known to be an important parameter characterizing the morphological
appearance of jets with purely toroidal magnetic field configurations (where we follow the definitions from~\citet{leis05}). 
For such jets, the pioneering work by~\citet{komiss99} has shown that
Poynting flux dominated (high $\sigma$), strongly magnetized (high $\beta_r$)
jets may develop a sharp `nose cone', where highly magnetized plasma accumulates
beyond the terminal shock of the jet beam. In contrast, strong backflows and
therefore cocoon-dominated jets emerged when the kinetic energy flux is
dominant (low $\sigma$). 
In the~\citet{leis05} parameter study extending these results to
also purely poloidal field configurations, no `nose cones' were found for poloidal topologies. 

\begin{figure}
\FIG{
{\resizebox{\columnwidth}{!}{\includegraphics[angle=90]{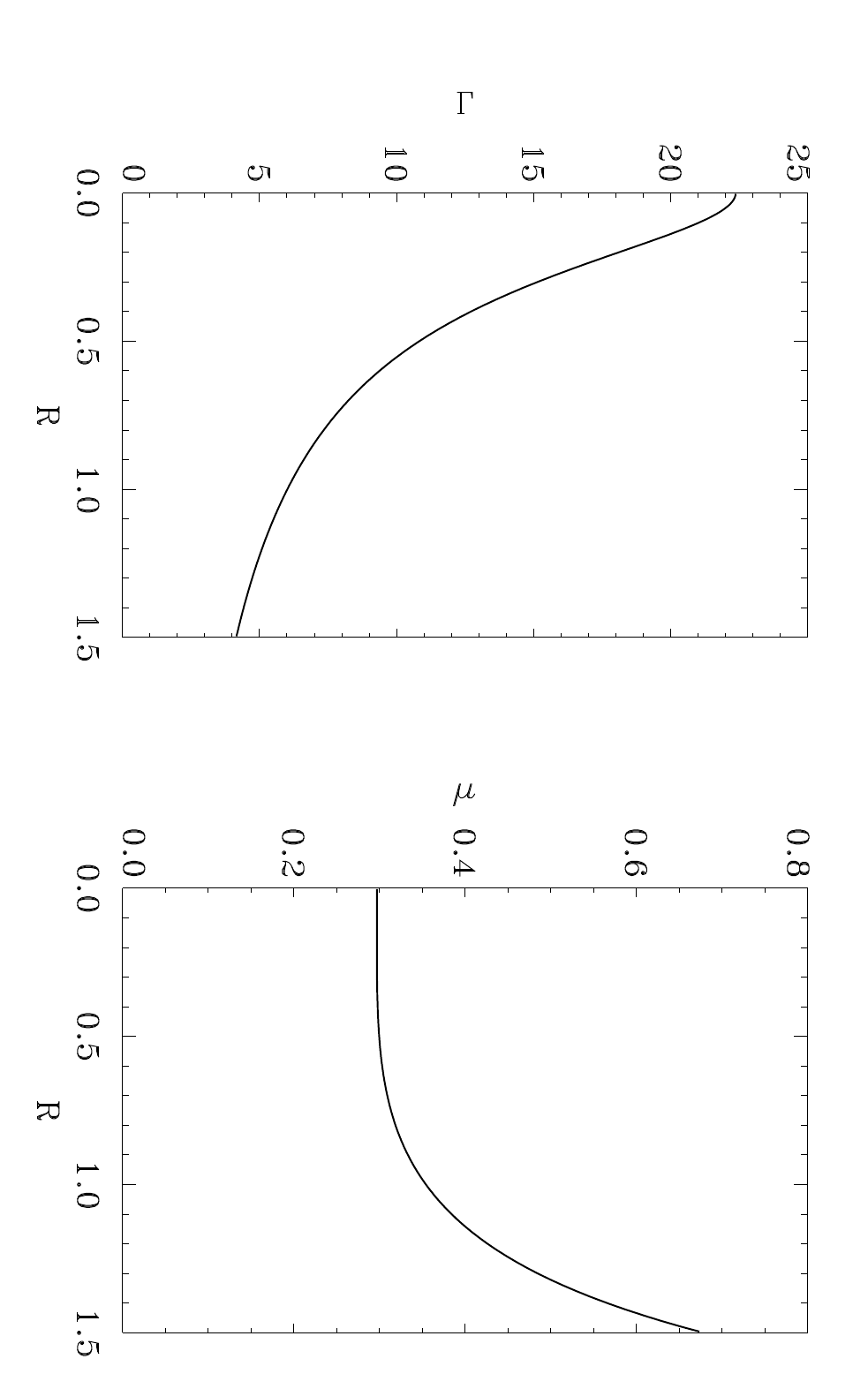}}}
}
\caption{Jet inlet profiles for the reference model Ref1. Left:
radial variation of the Lorentz factor $\Gamma$. Right: variation of the inverse pitch $\mu$.}\label{figlorprof}
\end{figure}

Due to the 2.5D nature of the equilibrium, all these parameters in fact vary mildly to strongly across the beam radius. In Table~\ref{tab2}, we computed beam-averaged values
\begin{equation}
 \bar{f}=\frac{1}{\pi R_{\rm j}^2} \int_0^{R_{\rm j}} 2 \pi R f dR . 
\end{equation}
From the values in this table, the sequence of models NR, MR, Ref1 with increasing $\alpha$ clearly has similar inlet conditions (low $\sigma$, equipartition like fields with $\beta_r\approx 0.3$), and explores the trend to $\bar{\Gamma}\approx 7$ flow regimes. In Figure~\ref{figlorprof}, we plot the actual variation of
the Lorentz factor $\Gamma$ across the inlet for the Ref1 model, and it is seen
that this model reaches axial flows with $\Gamma$ up to 22. The figure also
shows the inverse pitch $\mu$ for the same model.
The inverse pitch $\mu$ is a quantitative measure of the twist in the jet magnetic field and is defined from
\begin{equation}
 \mu = \frac{R_{\rm j} B_\varphi}{R B_Z}. \label{pitch}
\end{equation}
This profile is important for a stability analysis of helically magnetized
jet flows, and as we have axial values of this inverse pitch of order $0.3$ 
and as the current is distributed over the entire jet section 
\citep{Appletal99}, 
we may expect that non-axisymmetric current-driven instabilities are less likely
to play a role in the dynamics. 
However, a quantitative answer to stability with respect to non-axisymmetric perturbations (both of Kelvin-Helmholtz as well as current-driven type) is presently lacking for radially structured jets such as those introduced here. While
simplified `top-hat' profiles (i.e. layers of uniform, axially magnetized flows) can still be analyzed analytically
\citep{hardee07}, a numerical analysis of the linearized relativistic MHD equations is needed to predict 
3D effects which we artificially suppress by axisymmetry assumption.
Naturally, the sequence of models Pol, Tor, Ref2 with varying field structure
will be characterized by very different stability properties against non-axisymmetric modes. 

\begin{table}
\caption{Model input parameters, as appearing in the profiles quantifying the radial variation of primitive variables.}             
\label{tab1}      
\centering                          
\begin{tabular}{l c c c c}        
\hline\hline                 
 Model & $\rho_{\rm j}/\rho_{\rm c}$  & $B_0, B_c, B_1$ & $\alpha$ & $p_{\rm j}$ \\   
\hline                        
NR &   0.1 & 1, 0.01, 1  & 4.99  & 1 \\      
MR &   0.1 & 1, 0.01, 1  & 9.9  & 1 \\      
Ref1 &   0.1 & 1, 0.01, 1  & 9.99  & 1 \\      
\hline
Pol &   0.1 & 1, 0.01, 0.01  & 999.9  & 1 \\      
Tor &   0.1 & 0, 0, 1  & 9.99  & 1 \\      
Ref2 &   0.1 & 2, 0.01, 1  & 9.99  & 1 \\      
\hline
Ref3 &  0.2 & 1, 0.01, 1 & 9.99    & 1 \\ 
Ref4 &  0.1 & 1, 0.01, 1 & 9.99    & 2 \\ 
\hline                                   
\end{tabular}
\end{table}

\begin{table}
\caption{Model parameters, in terms of beam-averaged inlet values for
various dimensionless ratios.}             
\label{tab2}      
\centering                          
\begin{tabular}{l c c c c c}        
\hline\hline                 
 Model & $\overline{\Gamma}$  & $\overline{\beta_r}$ & $\overline{\sigma}$ & $\overline{M},\overline{\cal{M}}$ & $\overline{M_a}$ \\   
\hline                        
NR &   1.15 & 0.3015  & 0.0067  & 3.6, 4.1 & 6.6 \\      
MR &   4.79 & 0.2902  & 0.0064  & 7.1, 34.2 & 13.6 \\      
Ref1 &  6.92 & 0.2899  & 0.0064 & 7.2, 50.3 & 13.7  \\      
\hline
Pol &   7.78 & 0.2833  & 0.0063  & 7.2, 56.9 & 13.9 \\      
Tor &   6.92 & 0.0009  & 0.00002  & 7.8, 53.3 & 664.5 \\      
Ref2 &  6.92 & 0.9147  & 0.0250  & 6.2, 44.7 & 7.1 \\      
\hline
Ref3 &  6.92 & 0.2899   & 0.0064 & 7.2, 50.3 & 13.7 \\ 
Ref4 &  6.92 & 0.1514 & 0.0064   & 5.3, 36.5 & 13.9 \\ 
\hline                                   
\end{tabular}
\end{table}

\begin{figure}
\FIG{
{\resizebox{\columnwidth}{12cm}{\includegraphics{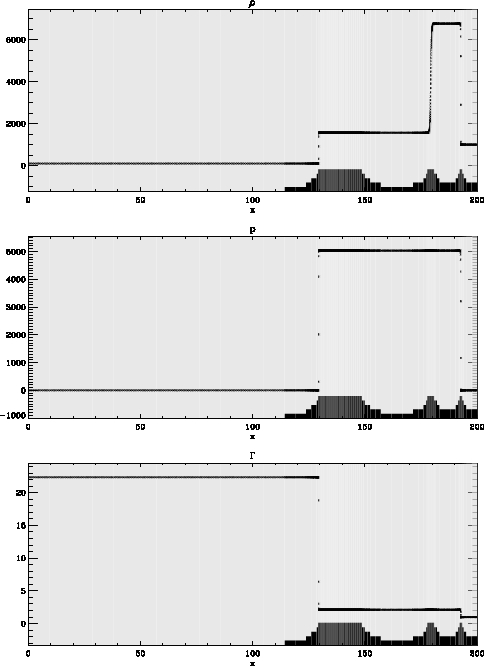}}}
}
\caption{A 1D Riemann problem indicative for the on-axis conditions of the reference jet. Shown is the proper density, pressure, and Lorentz factor,
at time $t\approx 200$.}\label{figRPA}
\end{figure}

\begin{figure}
\FIG{
{\resizebox{\columnwidth}{12cm}{\includegraphics{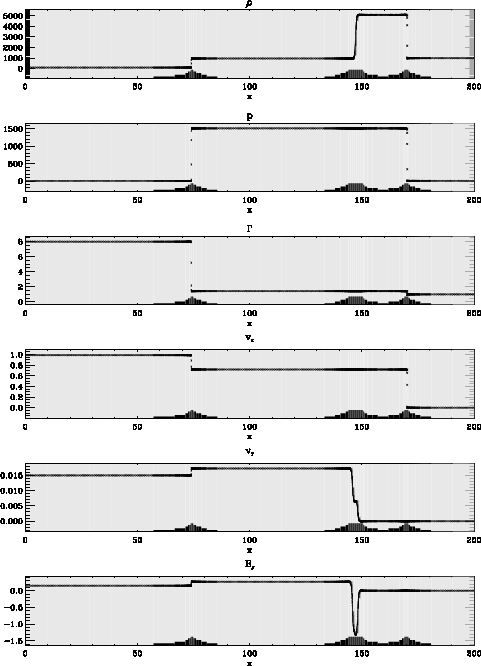}}}
}
\caption{A 1D Riemann problem indicative for $R\approx 0.75$ (midway) conditions of the reference jet. Shown is the density, pressure, Lorentz 
factor, and $v_x$, $v_y$ and $B_y$
at time $t\approx 200$.}\label{figRPB}
\end{figure}

As a means to anticipate the 2.5D results, and to verify whether the employed resolution suffices 
to capture the flow details, we first show the result of solving two Riemann problems that relate to the initial conditions in the reference case 
(Ref1). We solve these Riemann problems numerically using
essentially identical settings for grid refinement and resolution. The on-axis conditions, where Lorentz factor $\Gamma\approx 22$ conditions prevail, are as in the 1D problem where for constant $B_x=1$, left state L1 is adjacent to right state R1 given by
\begin{equation}
\begin{array}{lccccc}
 & \rho & p & v_x & v_y & B_y \\
\hline
L1:  & 100 & 1 & 0.999 & 0 & 0 \\
R1:  & 1000 & 1.5 & 0 & 0 & 0 \\
\hline
L2:  & 100 & 1 & 0.992 & 0.015 & 0.15 \\
R2:  & 1000 & 1.5 & 0 & 0 & 0 
\end{array} 
\end{equation}
States L2 and R2 give a Riemann problem that relates to the conditions at radius $R\approx 0.75$, midway through the jet. 
It must be noted that this relation is somewhat ad hoc, since the two-dimensional case has significant variation of $B_Z(Z)$ and $p(Z)$ at the head of the jet, which can not be mimicked in a 1D Riemann problem. Nevertheless, Figs.~\ref{figRPA}-\ref{figRPB} show the result at $t\approx 200$ for these Riemann problems. We capture in all cases both fast forward and reverse shocks very accurately, despite the fairly extreme parameters. 
The contact discontinuity is smeared out over many cells, but thanks to the AMR, adequately resolved. 
More problematic is the separation between contact discontinuity and the varying tangential field and velocity components in the Alfv\'en signals in the second Riemann problem. 
Still, their amplitude and variation is appropriately represented (and the entropy is constant through these discontinuities). These 1D tests qualify the shock-related pressure variations to be of 
several orders of magnitude, and indicate that we will be able to follow jet dynamics on a similar timescale within the computational domain.

\subsection{Jet morphologies: from non-relativistic to relativistic}

\begin{figure*}
\FIG{
{\resizebox{2.0\columnwidth}{8cm}{\includegraphics{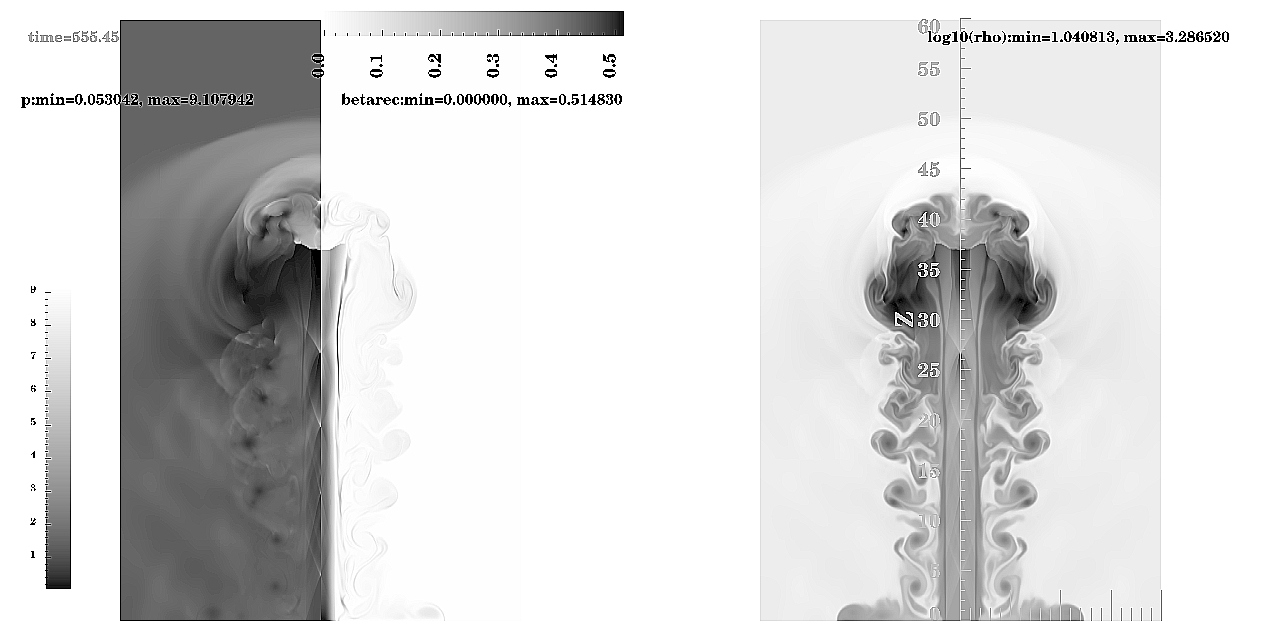}}}
}
\caption{For the low speed $\bar{\Gamma}\simeq 1.15$ jet (NR), we show the pressure and reciprocal plasma beta $\beta_r$
(left) and proper density (right) distribution at time $t\simeq 555$, corresponding to 185 light crossing times of the jet diameter.}\label{figNR1}
\end{figure*}

\begin{figure*}
\FIG{
{\resizebox{2.0\columnwidth}{12cm}{\includegraphics{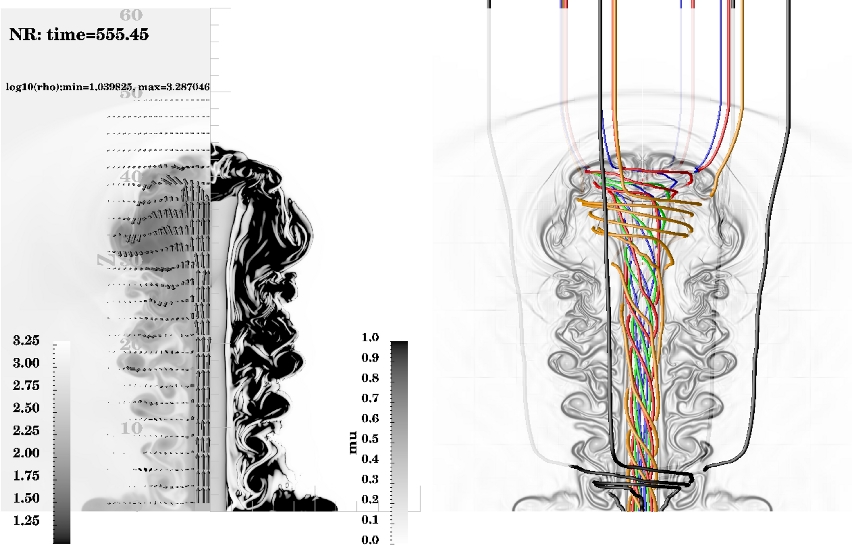}}}
{\resizebox{2.0\columnwidth}{12cm}{\includegraphics{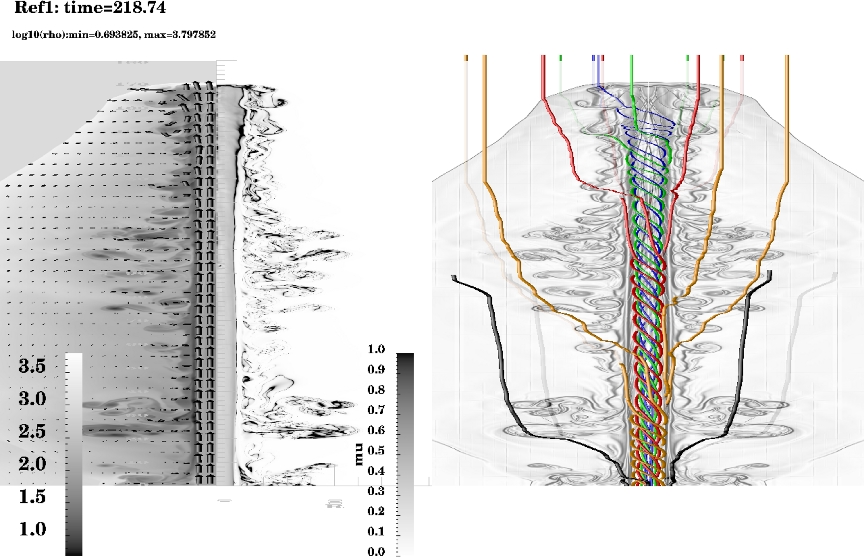}}}
}
\caption{For the NR jet from Fig.~\ref{figNR1} at time $t\simeq 555$ on domain $20\times 60$, the top panels show the flow
topology on top of density plot (leftmost panel), along with a quantification of
the inverse pitch throughout the jet: values above $\mu=1$ are colored black. In the right panel, a translucent Schlieren plot of the density is combined with field lines. The latter are colored according to their radial starting position
(progressivley more outwards: blue $R=0.3$, green $R=0.6$, red $R=0.9$, yellow $R=1.3$, black $R=1.8$).
The bottom panel gives the same info for the reference model Ref1, at a
different time $t\simeq 220$, using a very different aspect ratio of $20 \times 160$.
}\label{figbfield}
\end{figure*}

\begin{figure*}
\FIG{
{\resizebox{2\columnwidth}{8cm}{\includegraphics[angle=90]{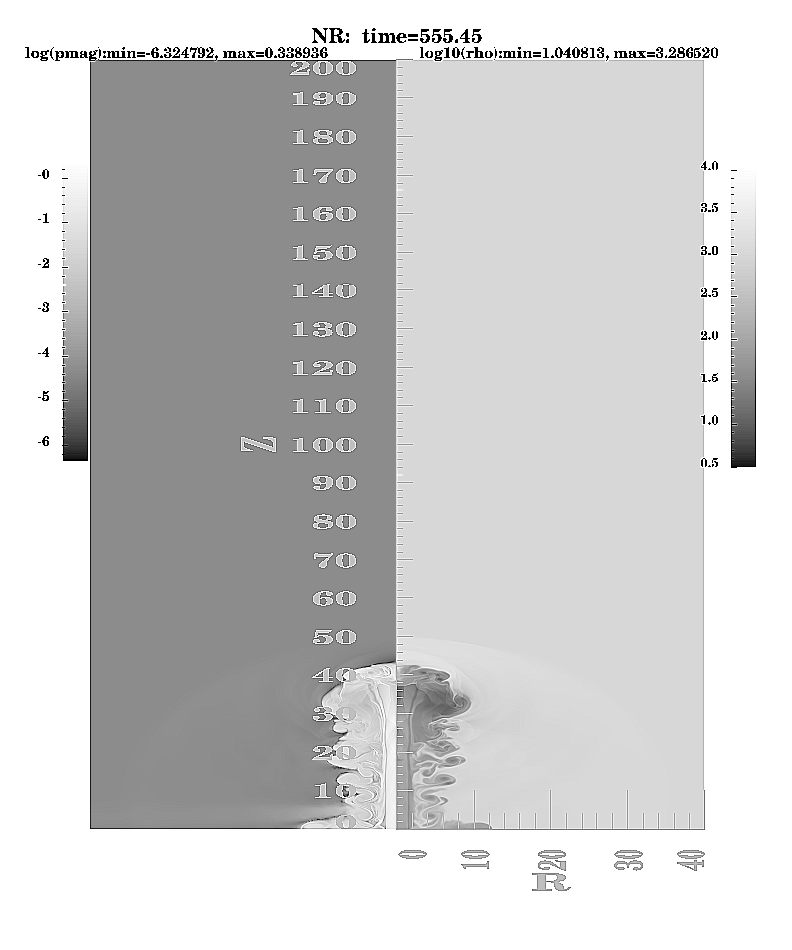}}}
{\resizebox{2\columnwidth}{8cm}{\includegraphics[angle=90]{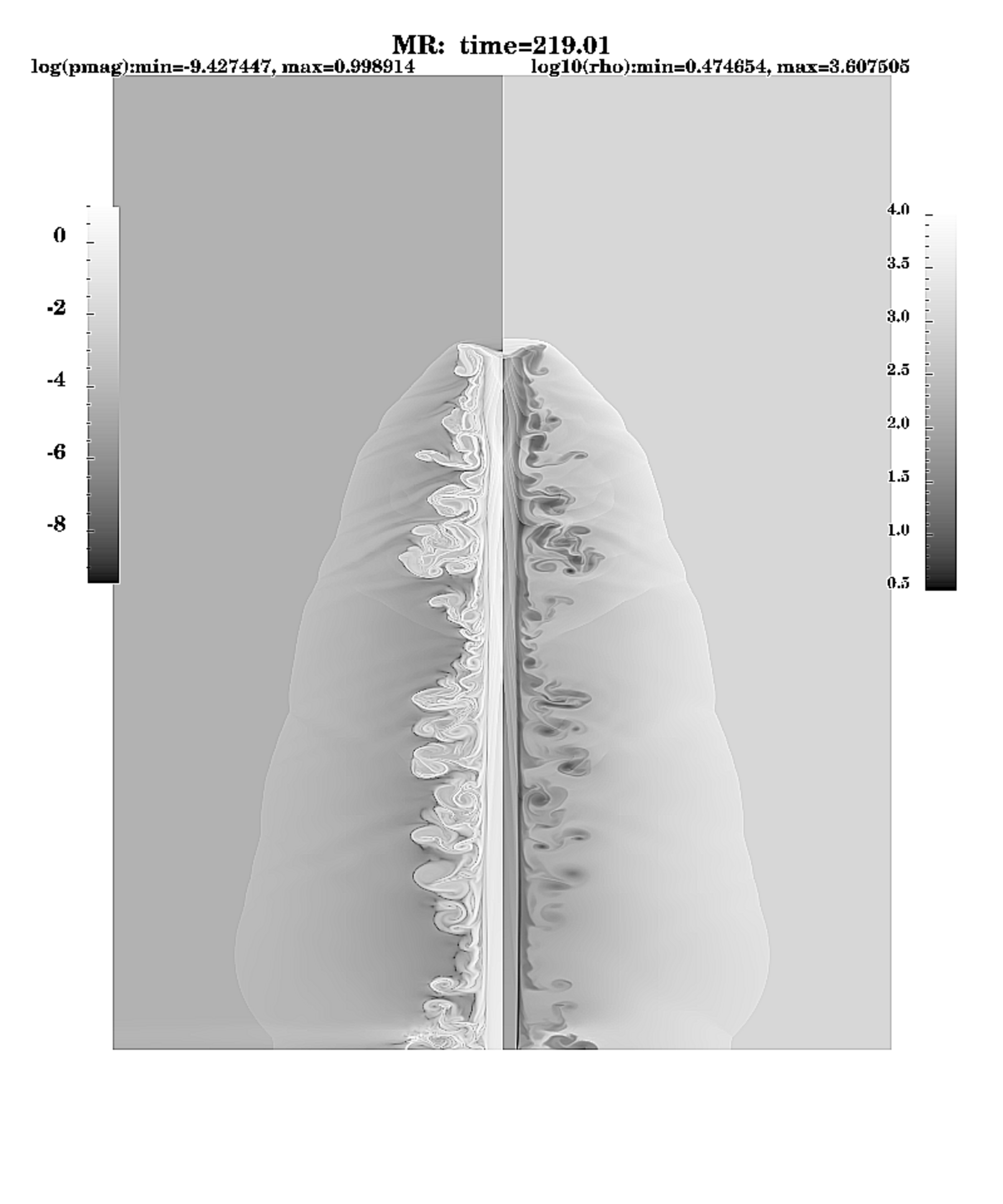}}}
{\resizebox{2\columnwidth}{8cm}{\includegraphics[angle=90]{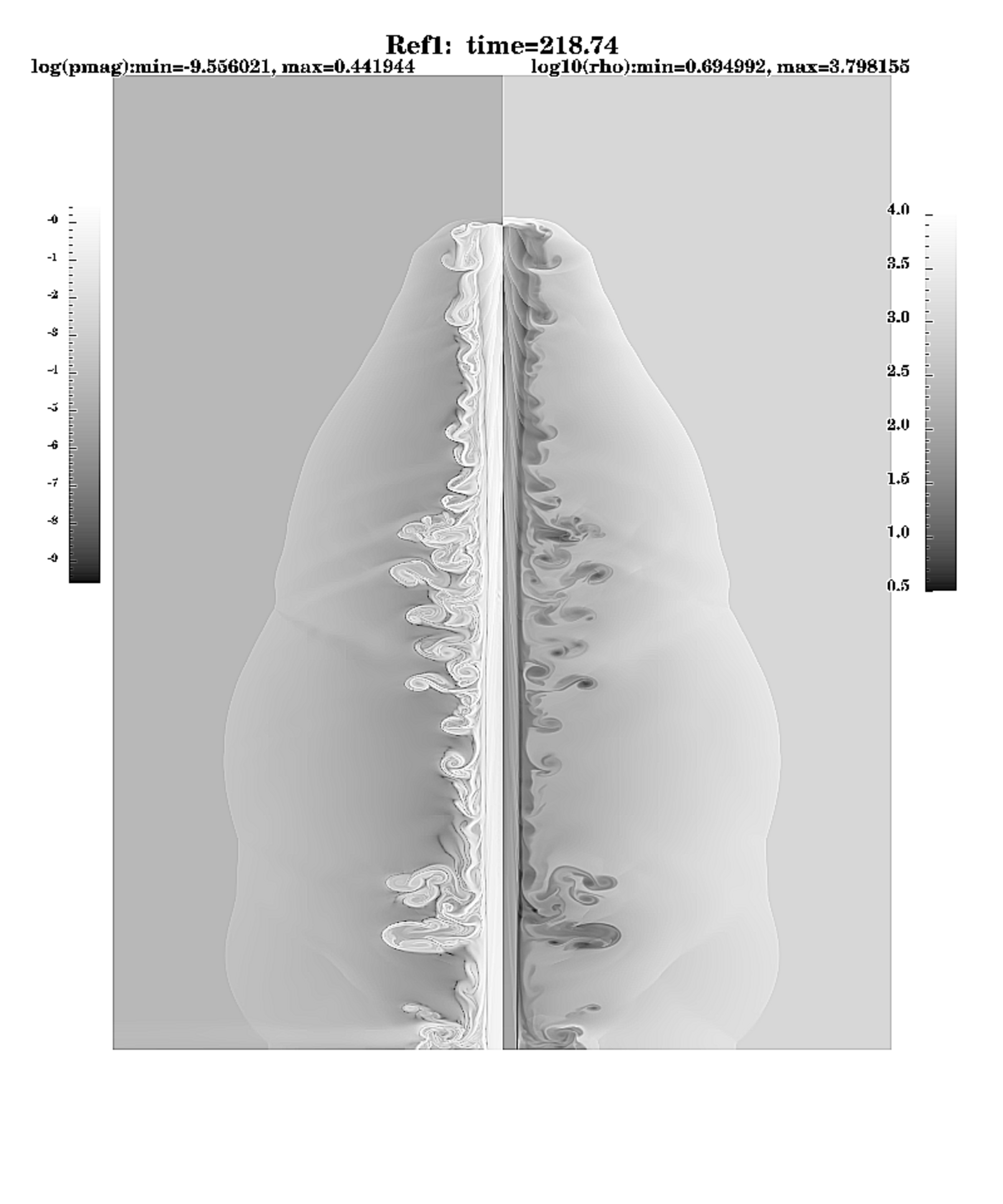}}}
}
\caption{The (logarithm of) magnetic pressure $p_{\rm mag}$ (bottom) and proper density $\rho$ (top) for models NR (at $t\simeq 555$), and MR and Ref1 (at $t\simeq 220$). The aspect ratio is identical in all frames, namely $40\times 200$.}\label{figNR}
\end{figure*}

\begin{figure*}
\FIG{
{\resizebox{\columnwidth}{6cm}{\includegraphics{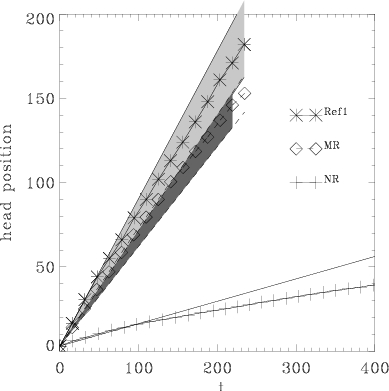}}}
{\resizebox{\columnwidth}{6cm}{\includegraphics{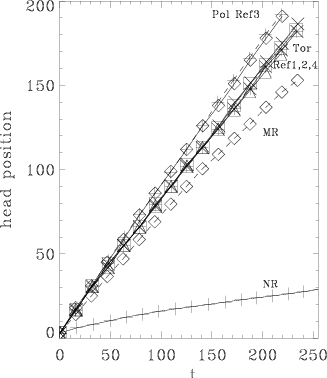}}}
}
\caption{The temporal evolution of the jet head position, for the sequence from non-relativistic, to the reference case (left panel),
and a comparison with an approximate formula for kinetic energy dominated jets.
At right, all simulated models are shown in a similar fashion, essentially demonstrating similar propagation speeds. 
}\label{figspeed}
\end{figure*}

\begin{figure*}
\FIG{
\begin{center}
{\resizebox{2\columnwidth}{7cm}{\includegraphics[angle=90]{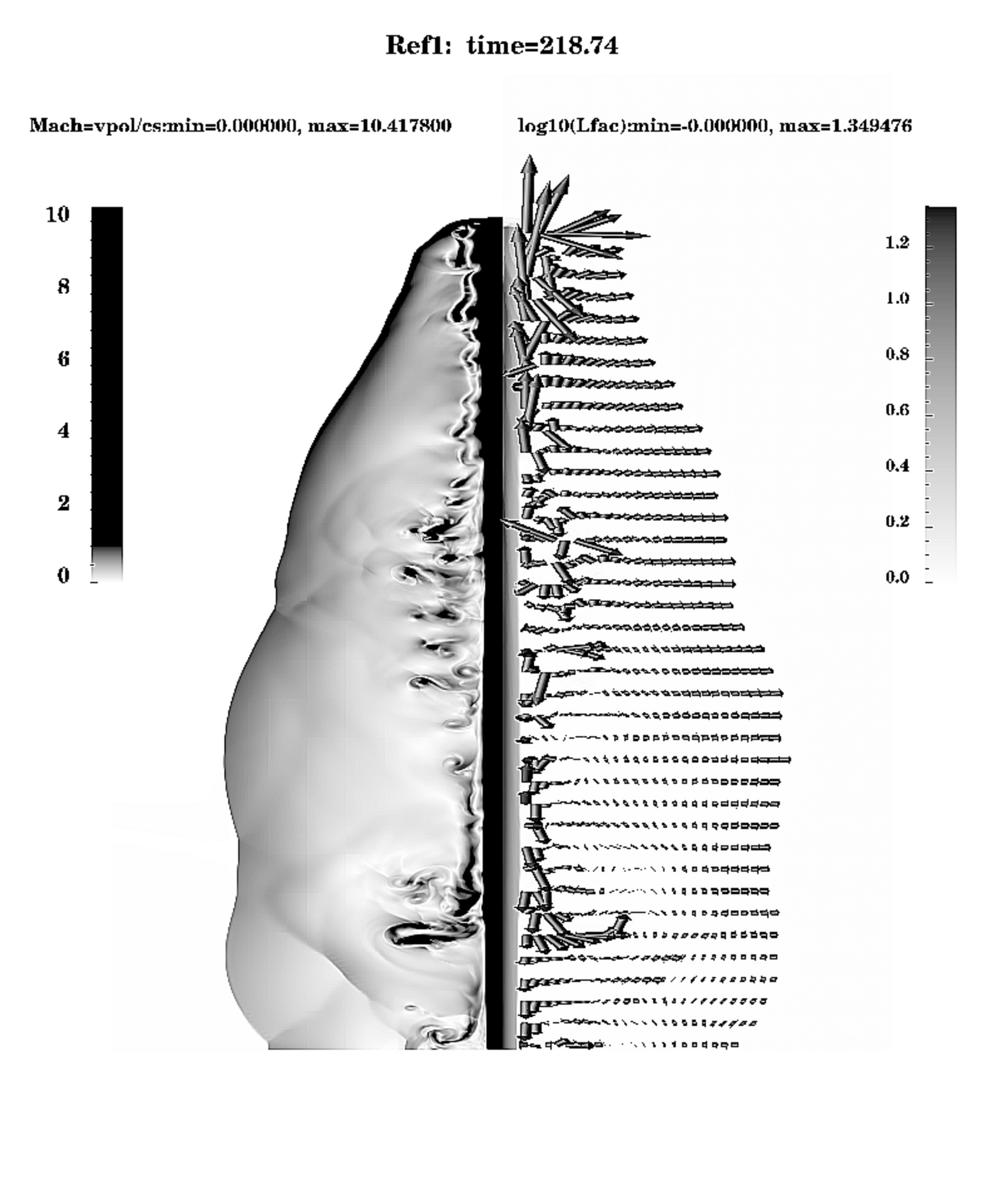}}}
{\resizebox{1.8\columnwidth}{3cm}{\includegraphics[angle=0]{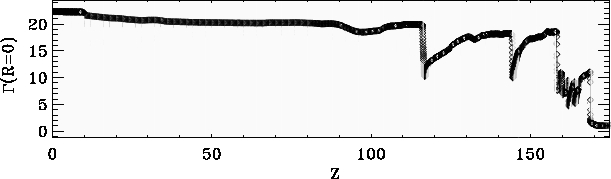}}}
{\resizebox{1.8\columnwidth}{3cm}{\includegraphics[angle=0]{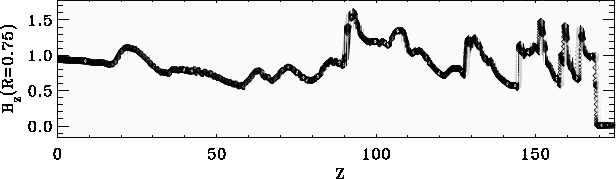}}}
{\resizebox{1.8\columnwidth}{3cm}{\includegraphics[angle=0]{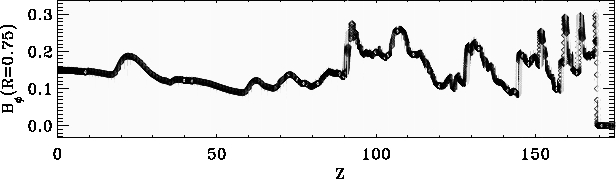}}}
\end{center}
}
\caption{For the reference Ref1 case, we visualize the flow pattern in the cocoon, along with a local Mach number quantification (bottom half of top panel).
Lower three panels: we show a cut along the symmetry axis of the Lorentz factor at the same time $t\simeq 220$, a cut of $B_R(Z)$ and $B_\varphi(Z)$ at the same time for $R=0.75$.}\label{figrefspeed}
\end{figure*}

To qualify the overall jet morphologies, we use the sequence of models 
characterized by increasing velocity (NR, MR, Ref1). In the near 
non-relativistic model NR, the jet beam terminates in a Mach disk, across
which a fair part of the directed kinetic energy is transferred to internal
energy. This gives rise to a prominent hot spot of high pressure
material, typically extending up to the contact discontinuity 
(or working surface) between shocked jet material, and shocked cloud material. 
The magnetic field is compressed by the reverse shock (Mach disk), 
enhancing the collimation efficiency which limits the sideways expansion of the shocked
beam matter (in between Mach disk and working surface). 
The shocked cloud material is bounded by an overarching bow shock. 
In this NR model, the inertia ratio between the jet and the external medium is 
low $\eta\sim\bar{\Gamma}_{\rm b}^2 \rho_{\rm b}/\rho_{\rm c} =0.132$, so that we expect 
a turbulent behavior near the working surface, which 
will lead to fairly strong disturbances to the jet. The compression of
the shocked external medium is moderate, and the sideways spreading of
shocked beam material is reflected by the external medium leading to
a bow shock and backflow.
In animated views of
the jet evolution, one can witness the formation of vortical patterns
at the contact interface in front of the Mach disk, which get pushed away laterally
and form complex backflow patterns surrounding the jet beam proper. In
doing so, recurrent cross-shocks are driven into the beam in the region behind
the terminal shock. These lead to interacting diagonal cross-shock
patterns internal to the beam, and they continually restructure the
shape and lateral extent of the Mach disk at the end of the beam. 
A plot at time $t\simeq 555$ of the pressure distribution (showing the high pressure hot spot), the reciprocal plasma parameter $\beta_r$,
and the proper density (in logarithmic scale) is shown in Fig.~\ref{figNR1}.

The magnetic field configuration for model NR at the same time $t=555$ is shown in the top panel of Fig.~\ref{figbfield}. In the beam,
slight helicity changes are associated with each diagonal cross-schock pattern,
with increased twist regions across a converging cross-shock, up to the
diverging fronts. In the cocoon region formed by the backflowing material,
rather strong azimuthal field components prevail, and the helically magnetized jet beam gets
surrounded by predominantly toroidal field regions. This can be seen in Fig.~\ref{figbfield} where the right panel gives a view of selected field lines, combined with a translucent surface containing a Schlieren plot of the rest frame density $\rho$. In the left panel, the velocity field shows the vortical motions associated with the backflows in the cocoon. The right half of this panel
maps the distribution of the absolute value of the inverse pitch $\mid \mu \mid$ from Eq.~\ref{pitch}. Regions
where $\mid \mu \mid$ exceeds unity are colored in black, and they give an indication of
the most strongly wound field regions. It is seen how the $\mu$ variation across
the inlet is rather preserved throughout the jet beam up to the Mach disk,
with modest variations associated with the diagonal cross-shocks as mentioned
above. Beyond the Mach disk and in the cocoon, predominant toroidal
field prevails, 
Hence, the reverse shock (Mach disk) compresses the toroidal magnetic field. 
There is also clear evidence for a more
poloidal, straight field layer in between the jet beam proper and the
backflow regions. 
These backflows push the poloidal magnetic field towards the axis,
and the resulting layer of strong poloidal magnetic field can be expected to increase the
lateral stability of the jet beam.

At all times in the evolution, only the jet beam, cocoon regions, 
and compressed region between beam and backflow, 
contain any significant magnetic pressure. This is shown in Fig.~\ref{figNR}, 
where we now contrast the three jet models NR, MR, and Ref1 near the end of 
the simulated time intervals on the entire computational domain. Note that the 
non-relativistic model is shown at $t\simeq 555$, while the faster jets MR and 
Ref1 are plotted at near identical, earlier times $t\simeq 220$. The bottom 
half of each panel quantifies the
magnetic pressure distribution $p_{\rm mag}$ (in a logarithmic scale), and
dynamically important values are found only up to the (turbulently deformed) 
contact interface between shocked beam plus
backflow material with shocked cloud matter. This figure also shows that the 
faster jet models have a clearly more elongated appearance, consistent with their
higher inertia $\propto \Gamma^2(\rho +\gamma p/(\gamma-1))$. The same figure 
also demonstrates that all our computations completely resolve the full bow 
shock pattern as they do not (yet) cross the lateral boundary. In figures
throughout this document, we will show appropriately scaled, zoomed in regions
of the domain only, to better illustrate flow details.
In Fig.~\ref{figbfield}, the lower panel shows the magnetic field and flow
information for the reference model Ref1, in the same manner as the top panel 
for the non-relativistic case NR. Note that times, vertical extent and aspect
ratio are quite different between the two cases, as evident from comparing them
with Fig.~\ref{figNR}. In the fast jet Ref1 (and also in the MR jet, which is
not shown in Fig.~\ref{figNR}), we again see that the magnetic field in the
jet beam roughly preserves the inlet helicity profile up to the Mach disk.
The beam is bordered by a near vertical field `sheet', and the vortical
patterns marked by lower density structures contain the most thightly wound
fields. Unlike the non-relativistic model though, the backflowing vortices
do not really 
form a rather extended cocoon surrounding the jet beam, but appear as more 
isolated narrow protrusions into the jet cavity bounded by the bow shock. The 
locations of the dominant toroidal field regions are therefore also 
rather localized. Thin strands of high field twist thus coincide with the 
lowest proper density spots which mark the centers of the vortical patterns.
As will be discussed later on, the rotational flow patterns in these vortices 
are locally supersonic.

In the preceding discussion, we contrasted flow morphology and dynamics for mildly relativistic to strongly relativistic cases
(NR to Ref1) all obtained in axisymmetric computations. All impulsively injected jet studies (where the jet gradually enters the simulation domain) tend to evolve from a `1D' to a `2D' phase once eddies start to dominate the cocoon dynamics and internal jet beam shocks fully develop. For the models discussed here, the first internal cross-shock has already clearly formed
at time $t\approx 15$. For the Ref1 case, multiple internal cross-shocks, and complex (ring-like) vortex sheddings
are prominent beyond times $t\approx 60$. A transition to a `3D' phase where jet disruption and/or significant
non-axisymmetric deformation occurs is precluded in our setup. According to the discussion on 2D versus 3D effects 
in \cite{carvalho}, jet disruption for classical HD jets may occur beyond distances given by $5 R_j M_j$, where $M_j$ indicates the jet Mach number.
If we take the jet averaged Mach number $\overline{M}$ from Table~\ref{tab2}, disruption to 3D can occur beyond time $t\approx 54$
and is likely to affect the further propagation. Using the more appropriate relativistic value $\overline{\cal{M}}\approx \Gamma \overline{M}$, this
disruption length still lies beyond the computed time interval though. 
Earlier 3D non-relativistic
hydrodynamic (e.g. \cite{bodo98}) and magnetohydrodynamic (e.g. \cite{batykep,oneill}) studies have been able to qualify some of the consequences of the axisymmetry 
restriction. 
\cite{bodo98} identified how small-scale structure emerges fast in full 3D hydro simulations of periodic
jet segments, and particularly light jets (just as those considered here) are rather prone to non-axisymmetric mode 
development. Dynamically important magnetic fields can help to stabilize the jet flow and mitigate the energy cascade to 
smaller scales. Classical 3D MHD simulations of periodic jet segments by \cite{batykep} identified how helically magnetized
jets can maintain jet coherency despite additional global non-axisymmetric instabilities. \cite{oneill} investigated
propagation characteristics of helically magnetized, light jets in 3D, following their propagation for lengths exceeding 100 jet radii. The presence of an intricately structured 3D `shock-web complex' at the frontal part of the jet
was clearly seen in renderings of the 
compression rate, most prominent in the high Mach number jet propagating in uniform surroundings.  
In view of these results, it can be expected that full 3D relativistic simulations for the sequence NR to Ref1
will show profound differences from our axisymmetric computations in the later stages of the simulation.

\subsection{Propagation speeds}
A straightforward qualification of the jet dynamics 
is obtained from their propagation speeds. It was shown in relativistic hydro
\citep{marti97},
and later used for relativistic MHD jets \citep{leis05}, that an estimate for the head advance speed could be made as
\begin{equation}
v_{\rm head}=\frac{\sqrt{\xi_b/\xi_a}}{1+\sqrt{\xi_b/\xi_a}}v_Z \,
\label{martiest}
\end{equation}
in which the jet beam internal versus (static $\Gamma_a=1$) ambient medium
enthalpies $\xi/\Gamma^2$ appear. This relativistic hydro estimate assumes 
pressure-matched conditions, and essentially uses momentum balance in the 
shock frame. 
In our magnetized jets, all taken in the kinetic energy dominated regime 
($\bar{\sigma}\ll 1$ see Table~\ref{tab2}),
we can expect a reasonable agreement with this formula.
Due to the internal beam equilibrium profiles, we can
use this expression to get a `lower' and `upper' bound for the expected
jet propagation. Using averaged beam profile values in Eq.~(\ref{martiest}), a reasonable
lower bound is obtained, while when specifying to beam axial values, a
kind of higher bound is found. In Figure~\ref{figspeed}, the left panel shows
the thus estimated speed ranges and compares them to the simulation results
for the sequence of low to high speed jets (NR, MR, and Ref1). For the slowest
jet model, both estimates essentially coincide, and this jet shows good
agreement with the predicted speed up to times $t\approx 100$, after which the jet starts to propagate slower. For the faster MR to Ref1 models, there is
also a trend to gradual deceleration, where the upper estimate prevails in
the first phase of the computation, but where we typically find actual propagation values in between the estimated bounds in the later stages. 
The right panel of Figure~\ref{figspeed} compares the computed propagation
velocities for all 8 jets. The fastest jets in this limited sample are
the nearly purely poloidal case, in accord with its higher beam 
(average as well as maximal, this time even going up to 70) Lorentz factor. 
In fact, this jet with (nearly) purely poloidal magnetic field behaves as a
hydrodynamic jet \citep{Majorama&Anile87} in first approximation.
Also the Ref3 case which has the reference jet structure penetrating a twice 
lighter external medium is fast, as it interacts weakly with the external 
medium. The propagation characteristics
of the other four models turn out to be very similar, and this again agrees
with their estimated values from Eq.~(\ref{martiest}). 

Still, prominent 2D effects
arise in our simulations, as e.g. seen in the shape of the bow shock in Fig.~\ref{figNR}.
This is influenced by the formation and interaction of the beam internal cross-shocks, as well as by the
complex evolution of the final Mach disk. The first cross-shock for the NR case forms near $Z\approx 4$, and
approximately remains at this location for times up to $t\approx 300$. For the MR case, the first cross-shocks
initially forms at about $Z\approx12$, gradually moving to larger distances. The reference Ref1 case initially forms
a strong cross-shock at about $Z\approx 25$, but its location and strength varies with time: at $t\approx 156$,
a fairly strong shock is visible at about $Z\approx 64$, but eventually a weaker shock remnant can be seen
in Fig.~\ref{figrefspeed} at about $Z=90$ (followed by a much stronger shock at $Z\approx 117$).
Despite these and other highly time-varying effects at the jet head, we find that
the high Mach number jets considered here thus agree well with 1D propagation estimates, consistent with
findings for non-relativistic, axisymmetric hydro jets presented in \cite{carvalho}. 

\begin{figure*}
\FIG{
{\resizebox{2\columnwidth}{12cm}{\includegraphics{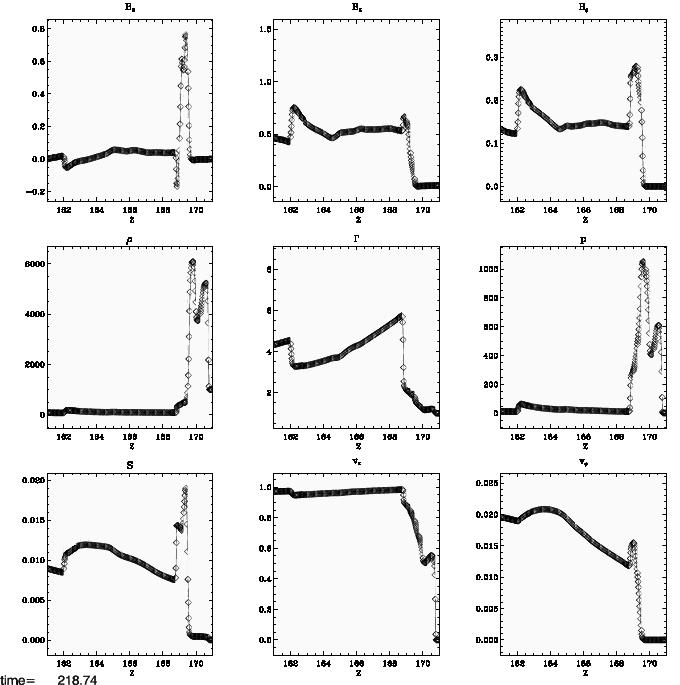}}}
}
\caption{For the reference jet Ref1, we show cuts through the jet head, taken at fixed radius $R=1$ and for $Z\in[161,171]$, and at the time $t\approx 220$. From left to right and top to bottom, we show
field components $B_R$, $B_Z$, $B_\varphi$, proper density $\rho$, Lorentz factor $\Gamma$, pressure $p$,
along with entropy $S$, and velocity components $v_Z$ and $v_\varphi$.}\label{figref1BB}
\end{figure*}

\begin{figure*}
\FIG{
{\resizebox{2\columnwidth}{6cm}{\includegraphics{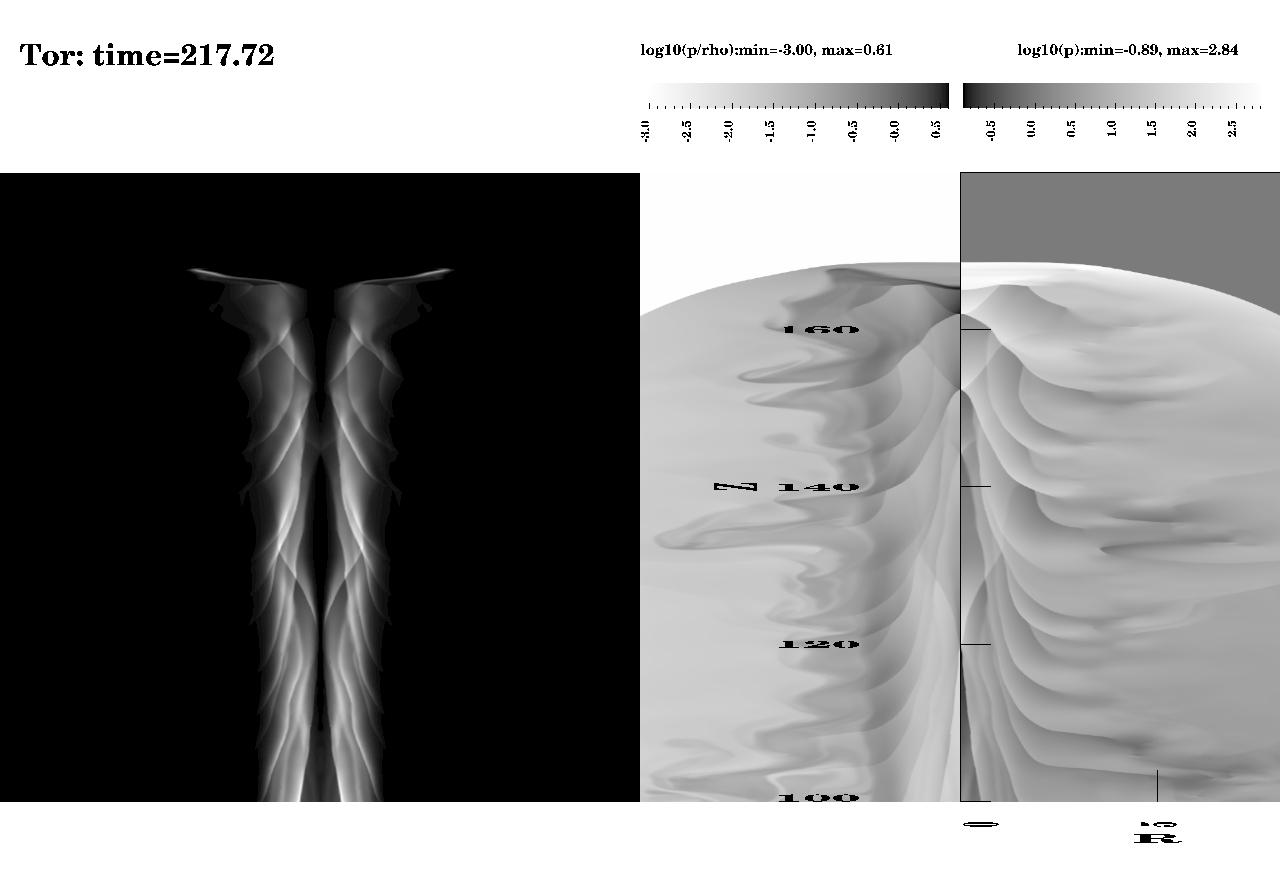}}}
{\resizebox{2\columnwidth}{6cm}{\includegraphics{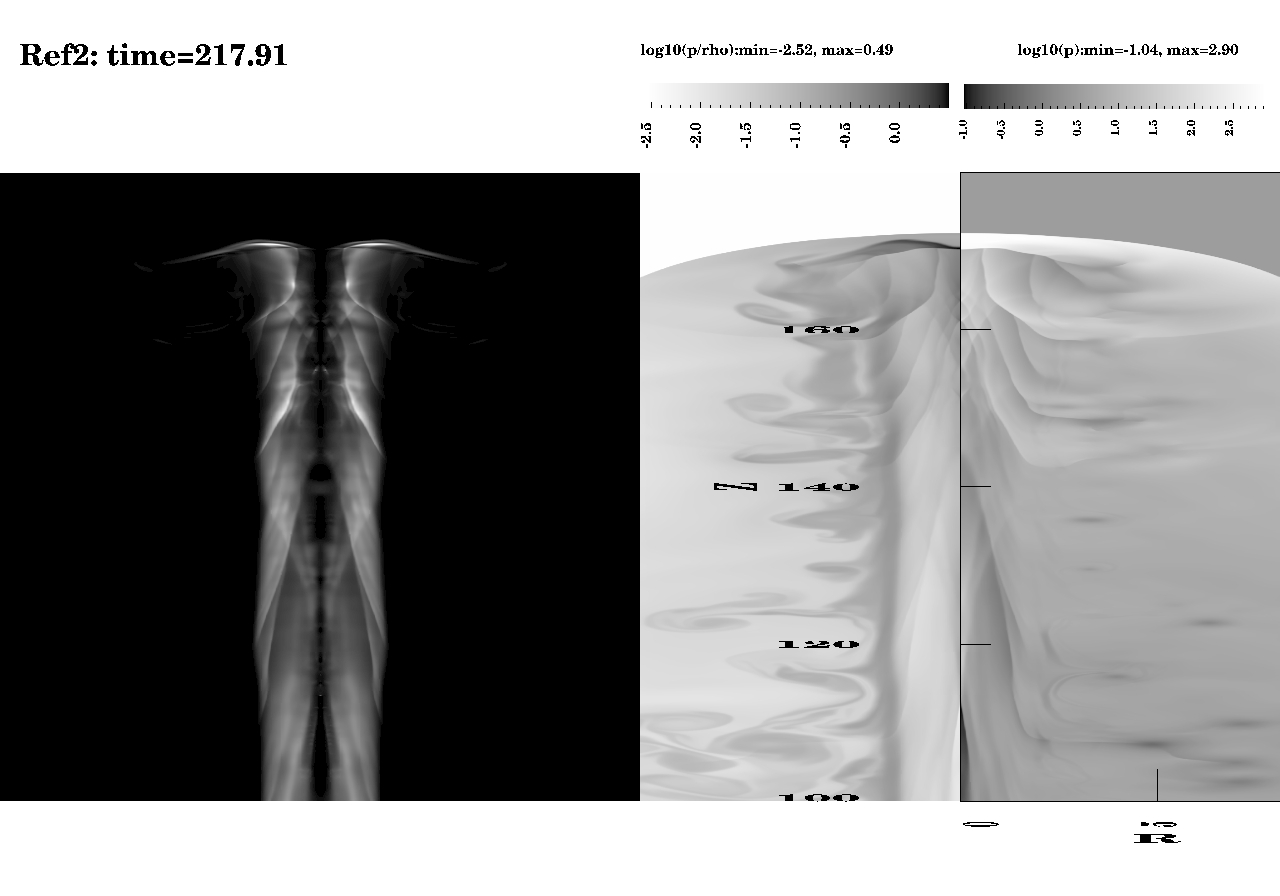}}}
{\resizebox{2\columnwidth}{6cm}{\includegraphics{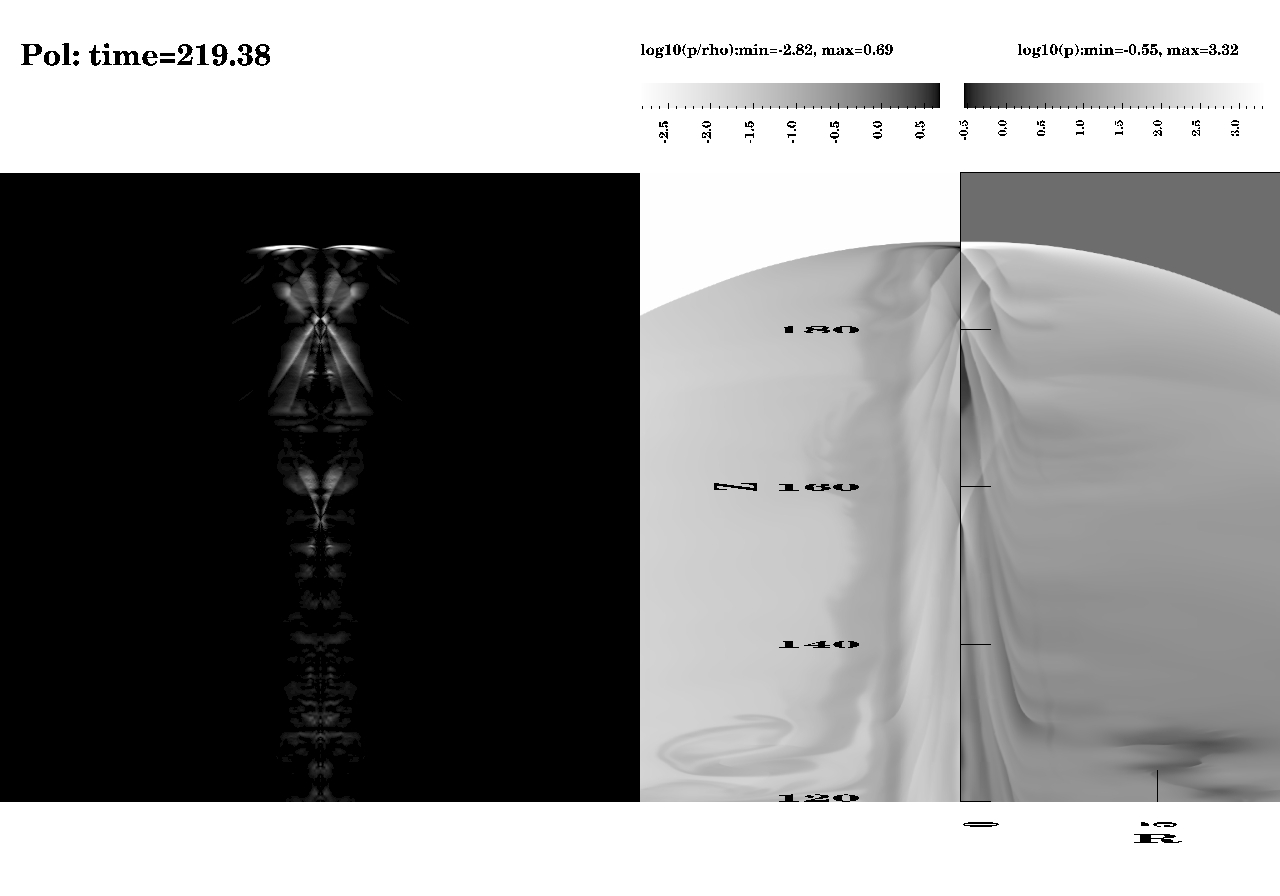}}}
}
\caption{For a sequence of models from toroidal (top), to a model with double the poloidal field magnitude as the reference case (middle),
to nearly purely poloidal magnetic field (bottom): maps of arbitrary scaled `power' given by $\Gamma^2 v^2 B^2 \sin^2(\Psi)$ (left), together with
logarithmic greyscale plots of the temperature and pressure distributions. All plots correspond to instantaneous
values at the endtime of the simulation, and only show a zoom on the top half of the entire domain.}\label{figseq3}
\end{figure*}

\begin{figure*}
\FIG{
{\resizebox{2\columnwidth}{8cm}{\includegraphics[angle=90]{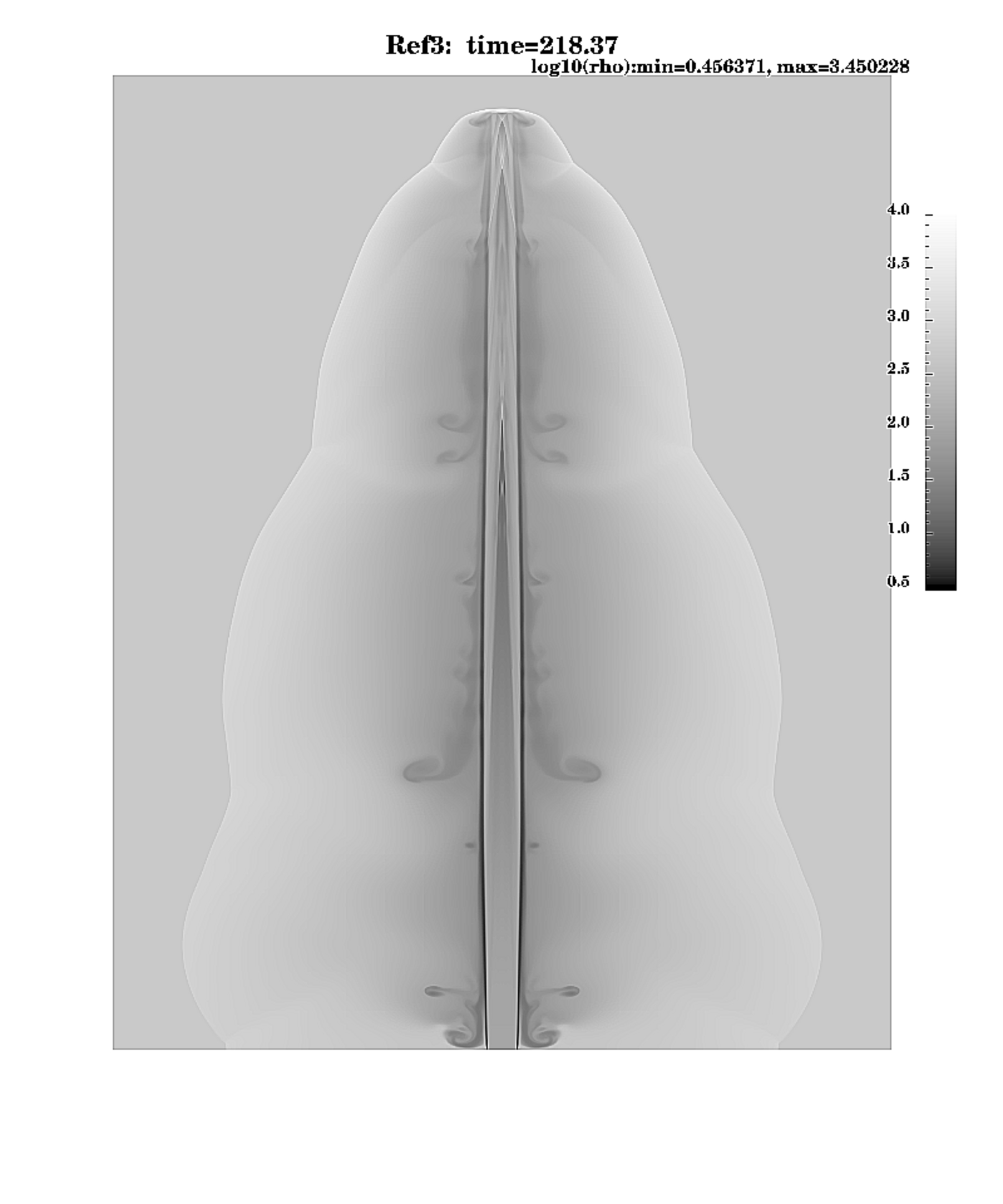}}}
{\resizebox{2\columnwidth}{8cm}{\includegraphics[angle=90]{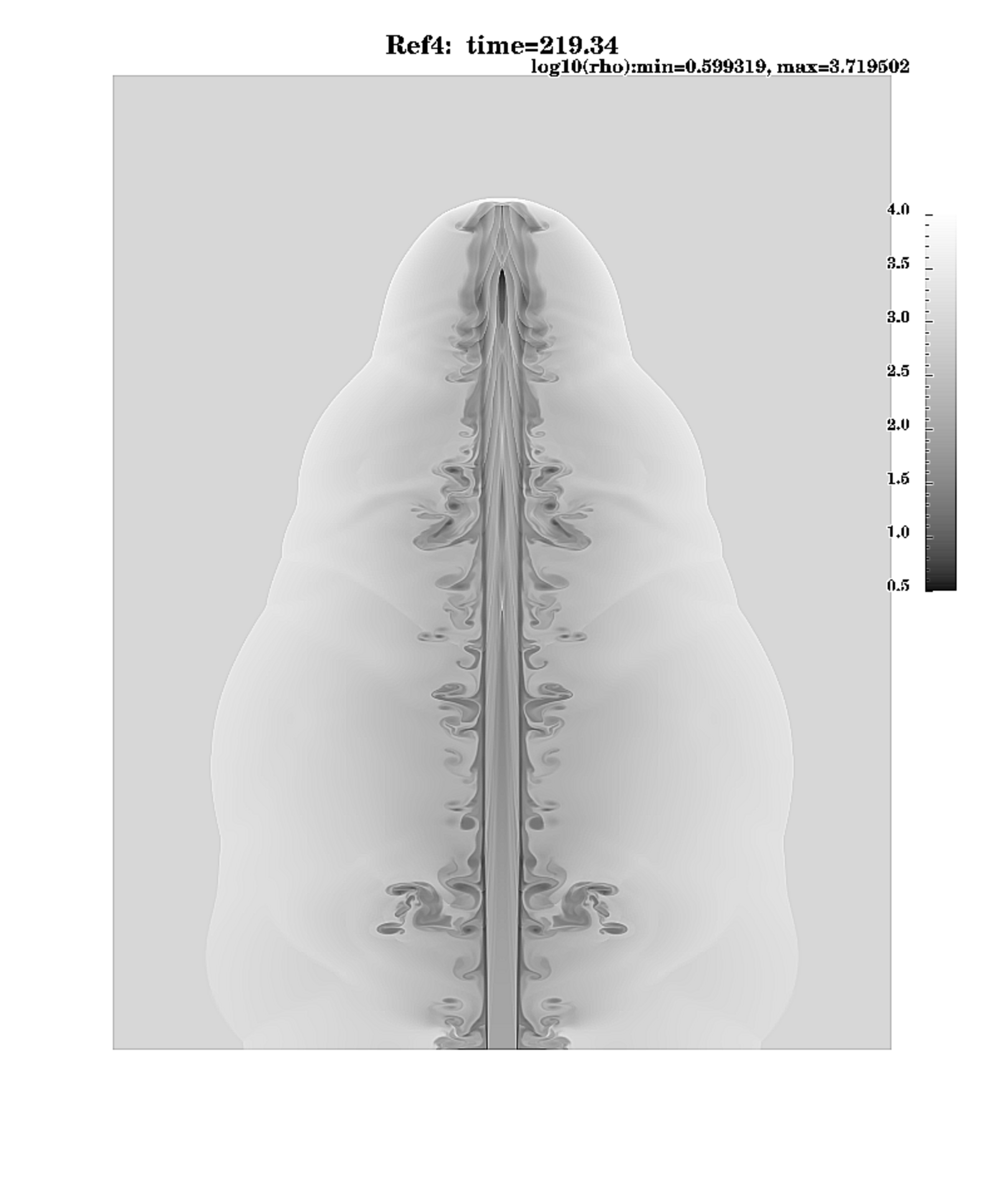}}}
}
\caption{For a heavier jet (top), versus a higher pressure jet (bottom): 
the proper density distribution at the endtime of the simulations. A reduced 
density contrast ensures fairly stable jet flows and little internal as well 
as cocoon structure. Increased jet pressure leads to more internal beam 
structure.}\label{figB}
\end{figure*}

In Fig.~\ref{figrefspeed}, we illustrate several properties of the flow field
for the Ref1 reference model. Using the same overall view as displayed in
Fig.~\ref{figNR}, the bottom half of the top panel indicates where regions of locally supersonic
poloidal velocities are encountered. All regions where $\sqrt{v_R^2+v_Z^2}/c_s$
exceed unity are colored black, and the remaining greyscale is used for the
subsonic range. The entire jet beam, together with the uppermost part of the
bounding bow shock is locally supersonic. 
The bow shock speed varies significantly along its entire boundary. As its speed is mostly transversal, we indeed confirm that its lateral supersonic expansion varies strongly behind the jet head 
as already found from axisymmetric hydro studies~\citep{carvalho}.
The upper half of this panel in Fig.~\ref{figrefspeed} shows the (logarithm of) the Lorentz factor in greyscale, and it is seen that only the jet beam up to
the Mach disk is seen to travel at a significant fraction of the speed of light
(up to on axis values $\Gamma\approx 22$, with some variation across the internal cross-shocks). The velocity vectors are
only drawn throughout the surrounding cavity, and show the complex
vortical patterns, the overal expansion of the cavity, and clear evidence of
localized regions of strong velocity shear. The latter aid in the formation of the turbulent structures by Kelvin-Helmholtz instabilities.
The lower panels in Fig.~\ref{figrefspeed} quantify the on-axis variation of the Lorentz factor at this endtime and the magnetic field variation midway the jet beam (at radius $R=0.75$). This shows that the beam internal cross-shocks act to convert some of
the directed kinetic energy, and due to slight variations of the cross-section of the beam,
matter accelerates again up to the next cross-shock. The magnetic field plays an important role in this
process, as the field strongly pinches the flow downstream of the shock. Note how in this reference case, Lorentz factor
$\Gamma\simeq 20$ flow occurs up to distance of $Z\simeq 158$, while
at the final Mach disk, the Lorentz factor is still above 10. In order to quantify the
deceleration process better, even longer simulations would be required. 

\section{Jet head structure and varying magnetic topology}

If we turn to the jet head structure in particular, we can quantify the compression
occuring at the Mach disk, and confront the field structure with that obtained in the
1D Riemann problem shown in Fig.~\ref{figRPB}. Fig.~\ref{figref1BB} shows the axial
variation at a radial distance $R=1$, zoomed in on the jet head structure (note the
limited $Z$-range). We discuss this structure from front to back.
Then, one first detects the forward shock (bow shock, at $Z\approx 170.8$) which leads to pronounced increases
in pressure and density, of a magnitude consistent with what the 1D problem demonstrated.
Another discontinuity of a pure thermodynamic nature trails at $Z\approx 169.9$. In a zone
$Z\in[169.6,169.9]$, we find the highest values for proper density $\rho\approx 6000$ and
pressure $p\approx 1000$. Magnetic
field variations all occur prior to $Z\approx 169.6$, where the density and pressure 
decrease from the values mentioned, to values still significantly above their beam values. 
We can detect distinct variation of all three magnetic field components within the region $Z\in [168.8,169.6]$, coinciding with 
increased entropy $S=p\rho^{-\gamma}$ within this zone. At the trailing shock (Mach disk)
located at $Z\approx 168.8$, the Lorentz factor drops from about 6 to 2. The discontinuity
seen at $Z\approx 162$ is the last internal shock prior to the jet head. There, as well
as at the Mach disk, the azimuthal field is enhanced, in accord with the increase in
transverse magnetic field seen in the 1D problem from Fig.~\ref{figRPB}. 
The analogy can not be expected to hold beyond such qualitative features: the actual 2D jet
shows significant multi-D variation: the magnetic field becomes nearly purely radial and azimuthal within 
a thin layer $Z\in [169.3,169.9]$, and for $Z>169.9$ the uniform vertical
cloud field of magnitude $B_c=0.01$ is found.

The sequence of models termed Pol, Ref1, Ref2, and Tor can be interpreted
as a series with similar inlet characteristics in terms of average Lorentz factor and Mach number, and changing magnetic topology from near poloidal to pure
toroidal. In accord with the decreasing influence of poloidal field in the
jet beam (from Pol to Tor), the backflows tend to show more fine structure with more eddies being
shed in the purely toroidal case. There is then accordingly more internal beam
structure: multiple cross-shocks are induced and interact. 
Once more, full 3D simulations are needed to
assess the generality of this result: the ringed vortex structures we obtain may break down and get stretched
in non-axisymmetric fashion. This can alter the overall field topology drastically and impact the backflow.
We find that in all but the
purely toroidal case, the regions of significant magnetic pressure are in the
beam and backflows. The purely toroidal case carried a weak magnetic field, 
and this can be seen to trace the region occupied by shocked beam matter
forming the inner cocoon.  

As a qualitative means to address how different magnetic topologies lead to varying jet morphologies we use this model sequence as follows. A formula quantifying the total emitted radiation by a single electron traveling at speed $\bfv$
in its relativistic
cyclotron motion about a magnetic field $\bfB$ is given by~\citep{Rybicki}
\begin{equation}
P_{\rm tot}=\frac{2 e^4}{12 \pi \epsilon_0 m_e^2 c} v^2 \Gamma^2 B^2 \sin^2(\Psi) \,\, ({\rm J s}^{-1})\,,
\label{power}
\end{equation}
where $\Psi$ indicates the pitch angle between particle velocity $\bfv$ and magnetic field $\bfB$ (in Tesla). 
In our relativistic MHD simulations, we treat the plasma as a single fluid characterized by its bulk speed $\bfv$. A crude means to infer
observational `synchrotron' intensity uses the bulk plasma speed in Eq.~(\ref{power}) together with the detailed knowledge on
the magnetic field distribution. In Fig.~\ref{figseq3}, the left panels produce arbitrary scaled maps of the essential local dependence
$v^2 \Gamma^2 B^2 \sin^2(\Psi)$ for the jet sequence, at the endtime of the simulations.  
It is to be noted that we
present instantaneous local values of this `power map', which is not appropriate for quantifying true synchrotron emission
for these relativistic jets, as we e.g. do not incorporate time retardation effects. 
A procedure to incorporate all relevant relativistic effects uses the output of relativistic HD simulations and solves the
relativistic radiative transfer equation assuming optically thin conditions, as done in \cite{komiss97}. Recently,
\cite{zakamska} used analytic self-similar axisymmetric RMHD jet models, to quantify both synchrotron emission as well
as polarization maps by performing the proper integration along the line of sight in the observer's reference frame.
In this paper, we restrict ourselves to quantifying the `power' dependence from Eq.~(\ref{power}), since for practical
purposes one would need to perform the data processing during the simulation, and make reasonable assumptions about the relativistic particle distribution function (which is not contained in the ideal RMHD simulation). The
local instantaneous values give quantitative information on the overall flow and field topology, and it is this aspect
which concerns us here mostly.
Only about the front half of the simulated domain is
shown in Fig.~\ref{figseq3}, and a simultaneous quantification of the pressure and `temperature' $T=p/\rho$ is shown at right. The highest pressure and temperature regions
coincide with the (very narrow) regions between the Mach disk and the contact surface, and show up as bright regions ahead of the 
jet beams. Consistent with the higher pressure found in the frontal compressed beam matter when more poloidal field is present, the associated (narrow)
hot spot eventually dominates.
In all cases though, the complex shock interactions seen in the beam clearly show up. Given the aforementioned trend towards 
more complex shock interactions when more toroidal field structure is present, the power maps reflect this trend directly. The angle-dependent factor in formula~(\ref{power}), together with the magnetic field variation with radius, also explains the trend seen from more outer beam sensitivity 
in the almost toroidal case, to pronounced inner beam sensitivity at diagonal cross-shock fronts for more poloidal field configurations.
This hints at the possibility to infer field topology characteristics from true synchrotron emission maps, which need to be constructed rigorously from simulations such as these presented here.

\section{Thermodynamic variations}
The final two models are shown in Figure~\ref{figB}, showing the logarithm of their proper densities. The top panel is for the case with reduced density contrast, where the jet penetrates less dense material than in the reference case Ref1.
The magnetized jet travels correspondingly faster, and a remarkably stable
beam with little backflow features results. The stand-off distance to the
first internal beam cross-shock is also increased. In contrast, the Ref4
model which has a higher overall pressure compared to the reference case,
shows a similarly rich pattern of vortex structures. Since the magnetic field
is identical in these models, the jet appearance seems to relate most
dramatically on density differences, at least for these near-equipartition,
kinetically dominated jets.
Comparing the increased pressure jet case with the reference case in Figure~\ref{figNR}, the richly structured backflows are slightly more pronounced, and the internal beam density contrasts are more evident. Within the simulated time, the higher pressure jet propagation has had about one more
phase of slight deceleration, reformation of the Mach disk, and consecutive acceleration shaping the overall bowshock. This
is in accord with its shorter internal sound crossing time. 
Quantitatively speaking, the highest proper density values (always found in the front
of the jet head) are found for the reference case Ref1 where its instantaneous maximum is found $\rho_{\rm max}=6280$ (see Fig.~\ref{figNR}), 
while the higher pressure case demonstrates values up to $\rho_{\rm max}=5236$.
For the lower density contrast, we find, as expected due to the lesser contrast and cloud density, a reduced value $\rho_{\rm max}=2818$.

\section{Outlook}

We studied the morphology and propagation characteristics of a series of
highly relativistic, helically magnetized jets. All were kinetic energy
dominated, and apart from the series of increasing averaged beam velocity,
were similar in propagation speeds and overall magnetization
(equipartition). With modest changes in magnetic topology, as well as in
internal pressure and density, fairly distinct differences can be detected
in the distribution of local power, as in the cocoon and internal
beam structure. The radially stratified jets studied here show significant
variation of Lorentz factor across their diameter, and the high central
`spine' (exceeding $\Gamma=20$) results in fairly elongated bow shocks. 
The magnetic helicity changes at internal cross-shocks act to reaccelerate the jet repeatedly.
The lighter jets show more fine structure in their backflows, and this will
likely continue over larger density contrasts than those studied here.
Previous studies \citep{leis05,komiss99} considered density ratios of 100
or more. The models here only considered density ratios of 5 to 10, and investigate jet propagation in
less dense environments which are known to lead to more stable jet configurations over longer distances.
The sheet of more poloidal field surrounding the jet beam proper which we found in the simulations 
also aids in stabilizing the beam.

In our simulations, the resulting jet propagation exhibits a small
deceleration of the head of the (still) relativistic jet along its axis, only seen in
the decrease in Lorentz factor along the axis occuring at the internal shocks. 
Such deceleration has been observed in various FR II jets, in correlation with an
increase of the 
magnetic field intensity and particle density \citep{Georg04,Sambrunaetal06b}. 
\citet{Tavecc06} have shown
that this deceleration is compatible with the entrainment of the external gas
by the jet. We plan to perform longer term simulations, in order to 
obtain a better description of the jet braking as well as a study of the impact of the external mass 
density compared to the jet one. In this kind of computation, we will explore regimes where the jet kinetic energy and the jet electromagnetic energy are of the same order. The use of the AMR strategy will 
be a powerful tool in the study of the associated magnetic field and density amplification 
occurring near the jet head. We then aim to provide statements from
modeling results, about the ratio of the jet density to the external medium density, and
use it to constrain better the observational values.

We also intend to explore the computationally challenging regime of Poynting
flux dominated jets with helical field topologies in future work. Since
most AGN jets are likely associated with high Poynting flux jets near the
source, it will be of interest to study the transition from Poynting to
kinetic energy dominated jets in even larger scale computations. Also, the
stability of these jets in fully three-dimensional simulations is as yet
unexplored. We demonstrated that the inlet jet magnetic topology for the
sample studied here seems to be maintained over a significant distance, and
the helicity measure showed only strong toroidal field concentrations in
the localized vortices developing from the backflows. This was consistent
with a trend followed from slower `non-relativistic' jets,  where toroidal
field gets created across the Mach disk, and subsequently mixed into the
cocoon. Since the toroidal field concentrations for the reference helical
jet are mostly in localized, supersonically rotating vortices, their
tendency to induce kink deformations may be less. Extremely high resolution
(grid-adaptive) 3D studies are called for to investigate this issue
further.

\begin{acknowledgements}
Computations have been performed on the K.U.Leuven High performance computing cluster VIC. ZM and RK acknowledge financial support from the 
Netherlands Organization for Scientific Research, NWO grant 614.000.421, and from the FWO, grant G.0277.08. 
These computations form part of the LMCC efforts.
\end{acknowledgements}

\end{document}